\documentclass[
aps,%
prb,%
reprint, 
twocolumn,%
groupedaddress,%
superscriptaddress,%
amsfonts,%
footinbib,%
floatfix,%
longbibliography
]{revtex4-2}

\RequirePackage{amsmath,amssymb,bm}
\RequirePackage{graphicx}
\RequirePackage[usenames,dvipsnames]{xcolor}
\usepackage{float}

\usepackage{pdfpages}
\makeatletter\AtBeginDocument{\let\LS@rot\@undefined}\makeatother

\usepackage[T1]{fontenc}

\usepackage{enumitem}
\usepackage{verbatim}

\usepackage{hyperref}
\hypersetup{
    colorlinks=true,
    citecolor= blue,
    linkcolor= blue,
    filecolor=blue,      
    urlcolor= blue,
    pdfauthor = {Chuan Chen},
    pdfcreator = {\LaTeX\ and \flqq hyperref\frqq},
}

\usepackage[capitalise]{cleveref}

\graphicspath{ {./Figs/} }

\newcommand{\Fref}[1]{Fig.~\ref{#1}}
\newcommand{\Frefs}[1]{Figs.~\ref{#1}}
\newcommand{\Eqref}[1]{Eq.~\eqref{#1}}

\newcommand{\Appref}[1]{Appendix~\ref{#1}}

\newcommand{\bbZ}{\mathbb{Z}}
\newcommand{\bh}{\bm{h}}
\newcommand{\bk}{\bm{k}}
\newcommand{\br}{\bm{r}}
\newcommand{\bR}{\bm{R}}
\newcommand{\bdelta}{\bm{\delta}}
\newcommand{\bGamma}{\bm{\Gamma}}
\newcommand{\bM}{\bm{M}}
\newcommand{\bK}{\bm{K}}

\makeatletter
\newcommand*{\rom}[1]{\expandafter\@slowromancap\romannumeral #1@}
\makeatother

\begin{document}

\title{Anyon polarons as a window into competing phases of the Kitaev honeycomb model 
under a Zeeman field}

\author{Chuan Chen}
\email{chenc@lzu.edu.cn}
\affiliation{School of Physical Science and Technology,
Lanzhou University, Lanzhou 730000, China}
\affiliation{
Lanzhou Center for Theoretical Physics, 
Key Laboratory of Quantum Theory and Applications of MoE,
and Key Laboratory of Theoretical Physics of Gansu Province,
Lanzhou University, Lanzhou 730000, China
}

\author{Inti Sodemann Villadiego}
\email{sodemann@itp.uni-leipzig.de}
\affiliation{Institut f\"ur Theoretische Physik,
Universit\"at Leipzig,
04103 Leipzig, Germany
}

\begin{abstract}
We compute the spectra of anyon quasiparticles in all three superselection sectors of the Kitaev model (i.e., visons, fermions and bosons), perturbed by a Zeeman field away from its exactly solvable limit,
to gain insights on the competition of its non-abelian spin-liquid with other nearby phases,
such as the mysterious intermediate state observed in the antiferromagnetic model.
Both for the ferro- and antiferro-magnetic models we find that the fermions and visons become gapless
at nearly identical critical Zeeman couplings.
In the ferromagnetic model this is consistent with a direct transition into a polarized state.
In the anti-ferromagnetic model this implies that previous theories of the intermediate phase viewed as
a spin liquid with a different fermion Chern number are inadequate, as they presume that the vison gap does not close.
In the antiferromagnetic model we also find that a bosonic quasiparticle becomes gapless
at nearly the same critical field as the fermions and visons.
This boson carries the quantum numbers of an anti-ferromagnetic order parameter,
suggesting that the intermediate phase may possess spontaneously broken symmetry with this order.
\end{abstract}

\date{\today}

\maketitle

\section{Introduction}
In his seminal work, Kitaev~\cite{Kitaev2006} introduced an exactly solvable model of spins in a honeycomb lattice
(see \Fref{fig:schematic}(a)) whose ground state is a fermionic BCS wavefunction projected to enforce single-site
occupancy~\cite{Burnell2011-za}. Although differing in aspects such as symmetry, it can be viewed as an exact realization
of Anderson's visionary ideas of resonating valence bond spin-liquid states.
Following the pioneering work by Jackeli and Khaliullin~\cite{Jackeli2009},
it has been recognized that the Kitaev interaction is one of the largest terms in models relevant for
materials
such as Na$_2$IrO$_3$, Li$_2$IrO$_3$, and $\alpha$-RuCl$_3$~\cite{Hwan_Chun2015-fk,Takagi2019-ok,Trebst2022}.
However, the non-Kitaev interactions present in these materials lead to magnetic ordering,
and despite years of investigations~\cite{Chaloupka2010-xs,Rau2014-ao,Hermanns2018-vh,Takagi2019-ok,Trebst2022},
fully understanding their impact has remained challenging because they disrupt the exact solvability of the model.
In $\alpha$-RuCl$_3$, it has been observed that applying a magnetic field destroys the zig-zag antiferromagnet order,
leading to a regime with oscillations in the longitudinal thermal conductivity~\cite{Czajka2021,Bruin2022-sn,Zhang2023-ve}.
But its origin remains debated, with proposals ranging from a true or
proximate spin liquid state~\cite{Villadiego2021-hu}, to originating from phase transitions associated with
stacking faults~\cite{Bruin2022-sn,Zhang2023-ve,Xing2024-fu}~\footnote{There are also reports of a quantized thermal Hall conductivity in this
regime~\cite{Kasahara2018-or,Yokoi2021-ul} but this has not been observed in other experiments~\cite{Czajka2021,Lefrancois2022-dl,Czajka2023-da,Zhang2023-ve}
and remains experimentally contested.
Several other probes have also been interpreted~\cite{Banerjee2016-nn,Banerjee2017-hq,Banerjee2018-km,Balz2019-lk}
as indicative of proximity to spin liquid behavior in this intermediate field regime.}.

Theoretically even the Kitaev model with only the addition of a Zeeman field remains poorly understood.
Numerical studies have shown that the stability range of the Kitaev spin liquid (KSL) depends crucially on the sign of
Kitaev interaction~\cite{Zhu2018,Gohlke2018-kg,Hickey2019,Patel2019,Ronquillo2019,Zhu2024-ti,Wang2024-ku}.
For ferromagnetic (FM) Kitaev interaction ($K < 0$ in \Eqref{eq:Hamiltonian}),
the KSL persists only up to a small critical field $h_{c}^{\mathrm{FM}} \approx 0.03 |K|$~\cite{Zhu2018,Gohlke2018-kg,Hickey2019},
beyond which the system enters the spin polarized (SP) phase.
On the other hand, with anti-ferromagnetic (AFM) Kitaev interaction ($K > 0$ in \Eqref{eq:Hamiltonian}), the KSL survives up to
much larger fields $h_{c1}^\mathrm{AFM} \approx 0.45 K$~\cite{Zhu2018,Gohlke2018-kg,Hickey2019}.
Interestingly, before the system enters the SP phase at $h^\mathrm{AFM}_{c2} \approx 0.7 K$,
there exists an intermediate phase (IP) without magnetic order~\cite{Zhu2018,Gohlke2018-kg,Hickey2019,Patel2019,Ronquillo2019,Jiang2019-jx,Holdhusen2024-yl},
which appears to be a gapless state characterized by a significantly enhanced low-energy density of states and
quasi-long-range spin correlation~\cite{Hickey2019,Zhu2018,Ronquillo2019,Patel2019,Gohlke2018-kg,Zhu2024-ti,Wang2024-ku},
but fully understanding its nature has remained elusive.

One of our central goals is to shed light on the nature of the IP
by investigating the modification of the spectrum of the quasiparticles of Kitaev model induced by a Zeeman field, and determine how
and at which values they become gapless and drive phase transitions into new states. 
The KSL under small Zeeman fields hosts an Ising topological order (ITO)~\cite{Kitaev2006} with three different types of anyon quasiparticles:
local bosons (like magnons), fermions (spinons), and non-abelian $\bbZ_2$ vortices (visons). While several previous studies have tried to understand the proximate phases by investigating the phase transitions driven by fermions,
i.e., via changes of their Chern numbers~\cite{Zhang2022-tc,Jiang2020-mo,Yilmaz2022-vm,Ralko2020-dr,Das2023-vp,Das2024-my,Nasu2018-et},
these studies, have largely ignored the evolution of other anyons such as the visons.
Our study will precisely fill in this gap, by computing the dispersions of quasiparticles of all three anyon types for the ITO,
including single visons.

\begin{figure}
\centering
\includegraphics[width=0.49 \textwidth]{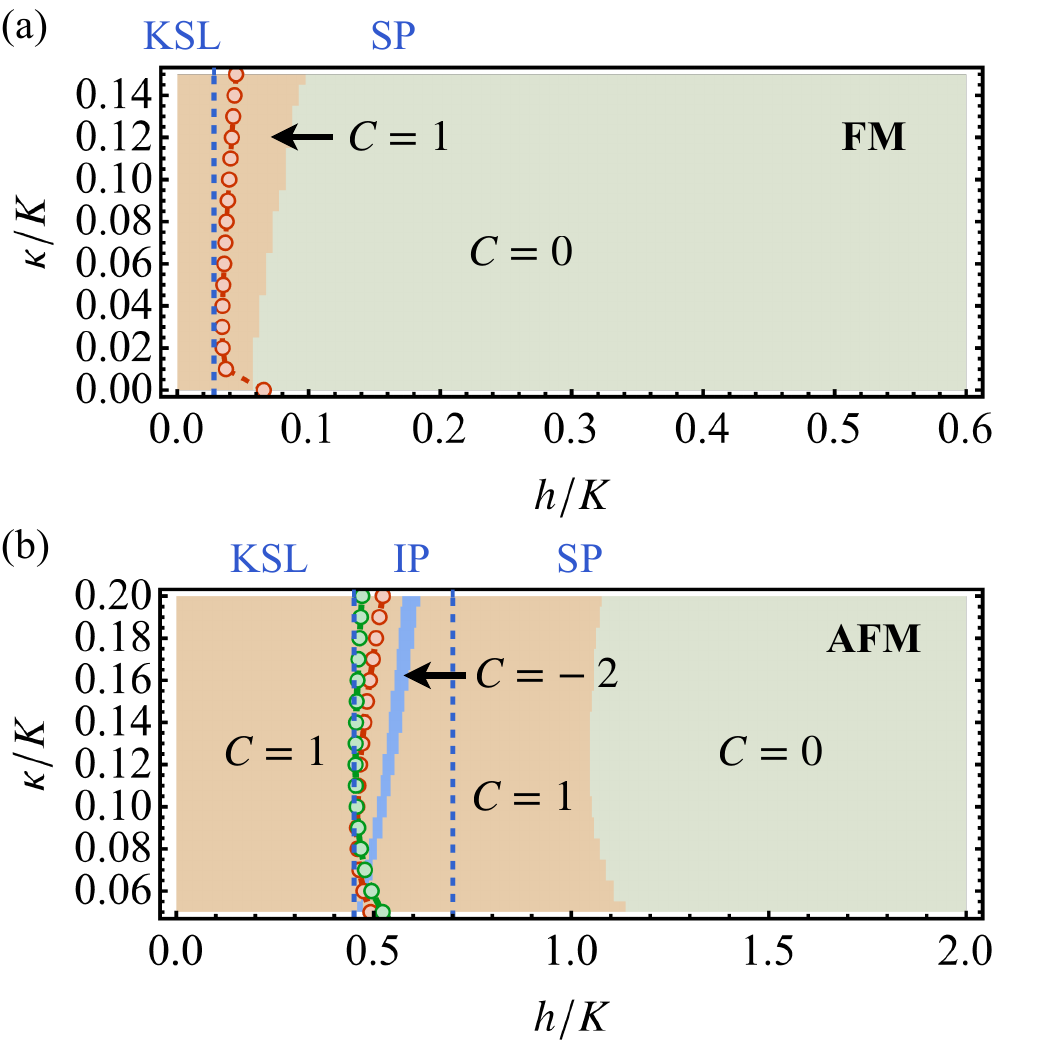}
\caption{
Fermion Chern number (color plot) and the critical fields indicating proliferation of single visons
(red dots) and bosonic vison pairs (green dots)
for (a) FM and (b) AFM Kitaev couplings.
The blue dashed lines indicate critical fields from numerical calculations (done at $\kappa = 0$)
from Refs.~\cite{Gohlke2018-kg,Zhu2024-ti}.
}\label{fig:phase-diagram}
\end{figure}

Interestingly, we will show that in both the FM and AFM KSLs, the single visons and fermions close their gaps at similar
Zeeman field values (see \Fref{fig:phase-diagram}).
The proximity of these distinct gap closings is remarkable, especially considering that our calculations are controlled only at small Zeeman fields.
Moreover, in the AFM case, we find that additionally a local bosonic quasiparticle (i.e., a dressed magnon) closes its gap at nearly the
same value as the fermions and single visons. Interestingly, we find that this boson transforms non-trivially under the model symmetries,
therefore its condensation is associated with spontaneous symmetry breaking, which we will argue likely has an in-plane 
AFM character.
The tantalizing proximity of the gap closings for these three distinct quasiparticles
is indicative of proximity to a highly non-trivial quantum critical point separating the AFM-KSL from the IP,
and the FM-KSL from the SP state, such as those proposed in Ref.~\cite{Zou2020-ni}.

\section{Model and preliminary considerations}
\begin{figure}
\centering
\includegraphics[width=0.46 \textwidth]{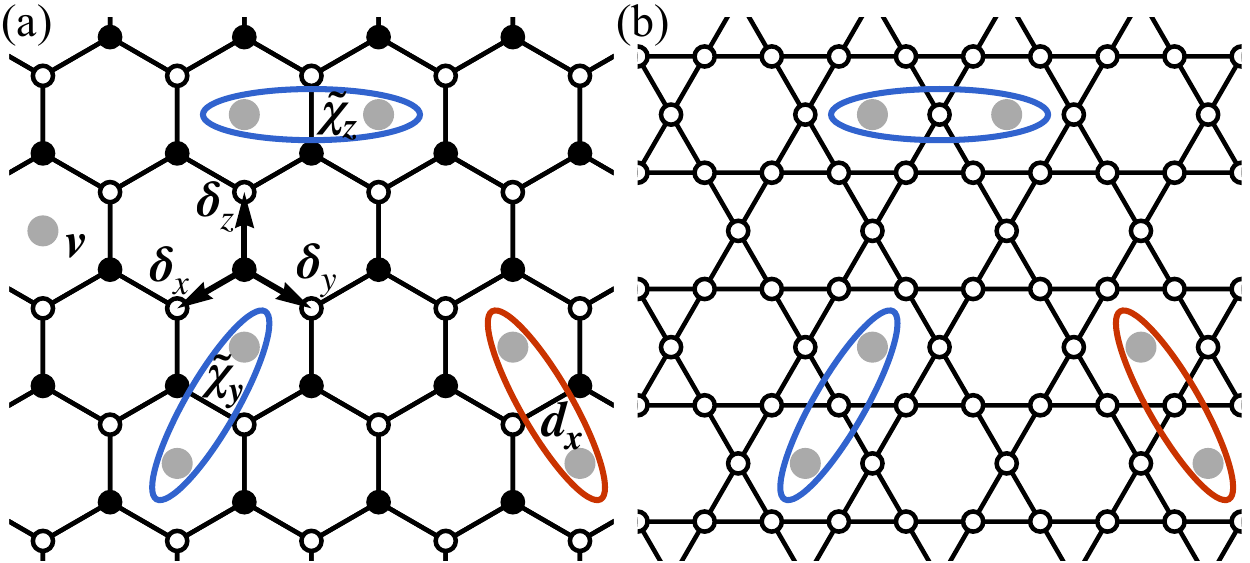}
\caption{
(a) Honeycomb lattice. Visons are centered in hexagons (gray dots).
 Nearest-neighbor visons can bind into pairs which are taken to  reside on the centers of links and thus move on a Kagome lattice (b).
These vison pairs can be fermionic ($\tilde{\chi}_\alpha$, blue ellipses)
or bosonic ($d_\alpha$, red ellipses).
}\label{fig:schematic}
\end{figure}

In order to understand the Kitaev model perturbed by a Zeeman field, 
it is convenient to also add explicitly the three-spin coupling ($\kappa$) that gives a
mass on Majorana fermions~\cite{Kitaev2006}.
Therefore, we investigate the following $K \mbox{-} \kappa \mbox{-} h$ model and treat the Zeeman term perturbatively:
\begin{equation} \label{eq:Hamiltonian}
    H = K \sum_{\br \in A, \alpha} \sigma_{\br}^\alpha \sigma_{\br+\bdelta_\alpha}^\alpha
    - \kappa \sum_{\langle j,k,l \rangle} \sigma_j^x \sigma_k^y \sigma_l^z
    - \sum_{\br} \bm{h} \cdot \bm{\sigma}_{\br}.
\end{equation}
%
For concreteness,
we will focus on the field along the $[111]$ direction: $\bh = h(1,1,1)$.
One key effect of the Zeeman field is the introduction of vison hoppings~\cite{Chen2023-vo,Zhu2024-ti} and the reduction of 
its excitation energy~\cite{Nasu2018-et,Patel2019,Hickey2019,Li2024-gb}, which tends to make them gappless.
In addition to individual visons, tightly-bounded vison pairs---residing on the lattice links labeled by $(\br,\alpha)$,
with $\br \in A$ and $\alpha = x,y,z$, see \Fref{fig:schematic} for an illustration---play an important role at low energies
due to attractive interactions between visons
(see \Appref{sec:E_two-vison}
for the two-vison excitation energy as a function of their separation)~\cite{Zhang2022-tc,Joy2024-cx}.
The fusion rule of the visons ($\sigma$) in KSL, $\sigma \times \sigma = 1 + \psi$, implies the existence of two types of low-energy vison pairs:
fermionic ($\psi$)~\cite{Zhang2022-tc} and bosonic vison pairs ($1$), with their respective annihilation operators denoted by
$\tilde{\chi}_{\br,\alpha}$ and $d_{\br,\alpha}$. See their detailed definitions in the following discussions.
Therefore, in this work, we investigate the $K - \kappa - h$ model by analyzing the behavior of 
these three types of anyon excitations:
i)~single visons (with annihilation operator $v_j$);
ii)~fermions, including $\tilde{\chi}_{\br,\alpha}$ and matter $c$-Majoranas;
iii)~bosons $d_{\br,\alpha}$.

\section{Visons}
In the absence of Zeeman coupling, the vison is immobile and has an energy $\Delta_v \approx 0.15 |K|$~\cite{Kitaev2006}. 
As discussed in Refs.~\cite{Chen2023-vo,Joy2022-ba}, to leading order in $h$, the effective single-vison Hamiltonian acquires a hopping term:
\begin{equation} \label{eq:H_vison}
    H_\mathrm{V} = \sum_{\langle i,j \rangle} ( t^{v}_{i,j} |v_i\rangle \langle v_j| + h.c. )
    + \sum_{i} \Delta_v |v_{i}\rangle \langle v_{i}|.
\end{equation}

\noindent Where $|v_i\rangle$ denotes the state in which a single vison resides at plaquette $i$ (an auxiliary second vison is kept immobile at a distant  point, see Ref.~\cite{Chen2023-vo} for details). As demonstrated in Refs.~\cite{Chen2023-vo,Chen2022-tr}, the FM and AFM Kitaev models 
have sharply distinct symmetry enriched topological orders
with respect to lattice translations (at non-zero $\kappa$). This leads for example to different ground state degeneracies of the two models in tori with odd numbers of unit cells: the AFM (FM) model has $1$ ($3$) ground state(s)
(a proof is provided in \Appref{sec:GS_parity}). 
This topological distinction also leads to sharply different vison properties.
In the FM model, visons have trivial non-projective translations and a larger hopping amplitude.
However, in the AFM model, visons have a projective translations with $\pi$-flux per unit cell and a much weaker hopping
which in fact vanishes in the limit of $\kappa \rightarrow 0$ to linear order in $h$~\cite{Chen2023-vo,Chen2022-tr,Joy2022-ba}.
However, since $\kappa$ is perturbatively generated to cubic order in $h$~\cite{Kitaev2006}, in order to capture the true behavior of the AFM model at finite $h$,
it is more accurate to take a non-zero $\kappa$ when $h$ is finite, this is why we have truncated the phase diagrams in
\Fref{fig:phase-diagram}(b) to display only small but finite $\kappa$ regions
(see \Appref{sec:pd-small-kappa} for more details).

The larger vison hopping in the FM case explains why their critical proliferation field, $h_v^{\mathrm{FM}}$,
is small (red dotted line in \Fref{fig:phase-diagram}(a)).
These visons close their gap at momentum $\bk = 0$, so the resulting phase is expected to preserve translational symmetry.
We obtain $h_v^{\mathrm{FM}} \approx 0.04 |K|$, which is in close agreement with the value $h_c^\mathrm{FM} \approx 0.03 |K|$~\cite{Gohlke2018-kg}
found in numerical studies for the transition into the SP state for the FM model.
On the other hand, for the AFM model, the critical vison proliferation field, $h_v^\mathrm{AFM}$, is an order of magnitude larger than
$h_v^\mathrm{FM}$.
The AFM vison closes their gap simultaneously at wave-vectors $\bk = \pm \bM/2$ (see \Fref{fig:vison-bands}).
This does not necessarily imply that the resulting state breaks lattice translations, because local order parameters are made from vison pairs,
and the momenta of these soft vison modes can combine into pairs with zero or finite momentum.
In fact, the analysis of boson vison pairs presented later indicates that the resulting state preserves the lattice translations.

\section{Fermions}
The fermionic vison pairs can be constructed as~\cite{Zhang2022-tc}:
\begin{equation} \label{eq:fermion-pair-def}
    \tilde{\chi}_{\br,\alpha}^\dagger |\Omega \rangle = \chi_{\br,\alpha}^\dagger |0; \Psi^c_0(\br,\alpha) \rangle.
\end{equation}
Here 
$\chi_{\br,\alpha}^\dagger = (b_{\br}^\alpha - i b_{\br+\bdelta_\alpha}^\alpha)$ creates a ``bare'' bond fermion formed by
the two $b^\alpha$-Majoranas on bond $(\br, \alpha)$.
$| 0; \Psi^c_0(\br,\alpha)\rangle$ is the ground state of a matter $c$-Majorana BdG Hamiltonian $H^c(\br,\alpha)$ with two $\pi$-fluxes
separated by the bond $(\br,\alpha)$, and has the same fermion parity as the true ground state with no visons, denoted by
$| \Omega \rangle$~\cite{Knolle2014-wl,Knolle2015-lx,Zhang2021-xi}. The Zeeman coupling can induce fermionic vison pairs' hopping and their hybridyzation with matter $c$-Majoranas.
The effective fermion Hamiltonian reads:
\begin{align} \label{eq:H_fermion}
    H_\mathrm{F} =& \sum_{\alpha \neq \beta} t^{\chi}_{\alpha,\beta} \left( \sum_{\br \in A} \tilde{\chi}_{\br,\alpha}^\dagger \tilde{\chi}_{\br,\beta}
    + \sum_{\br \in B} \tilde{\chi}_{\br-\bdelta_{\alpha},\alpha}^\dagger \tilde{\chi}_{\br-\bdelta_\beta,\beta} \right) \nonumber \\
    & + \sum_{\br \in A,\alpha} \sum_{\bR} \left[ p_{\bR,\alpha} (  -i c_{\br+\bR} - c_{\br+\bdelta_\alpha-\bR} ) \tilde{\chi}_{\br,\alpha} + h.c. \right]
    \nonumber \\
    & + \sum_{\br,\alpha} \Delta_\chi \, \tilde{\chi}_{\br,\alpha}^\dagger \tilde{\chi}_{\br,\alpha} + H^c(0).
\end{align}
Here $\bR$ is a Bravais vector, $H^c(0)$ is the usual matter $c$-Majorana BdG Hamiltonian of the $K-\kappa$ model with no visons from Ref.~\cite{Kitaev2006}.
$\tilde{\chi}_{\br,\alpha}$ can be viewed as residing in the center of a honeycomb bond $(\br,\alpha)$ and thus moves
on a Kagome lattice, as illustrated in \Fref{fig:schematic}(b).
There are $4$ complex fermions per unit cell and therefore $8$ BdG bands in \Eqref{eq:H_fermion}.
It can be shown also that $t^\chi_{\alpha,\beta} = (t^\chi_{\beta,\alpha})^*$ and $p_{\bR,\alpha} \in \mathbb{R}$.
Our approach to compute these effective couplings is very similar to that of Ref.~\cite{Zhang2022-tc}, 
but we have constructed $p_{\bR}$ from a polaronic picture which uses the \emph{fully dressed} wavefunctions
of the $c$-Majonara in the presence of the flux pair associated with $\chi_{\br,\alpha}$.
On the other hand, Ref.~\cite{Zhang2022-tc} used  bare plane-waves for the matter $c$-Majoranas without the fluxes.
The resulting $p_{\bR}$ in our case is highly localized, in contrast to the long-range form from Ref.~\cite{Zhang2022-tc}.
This difference does not affect significantly the value of $h$ for the first fermion gap-closing transition,
but changes the Chern number and width of the resulting phase (see a detailed comparison between our approach
and Ref.~\cite{Zhang2022-tc} in \Appref{sec:compare}).

\begin{figure}
\centering
\includegraphics[width=0.49 \textwidth]{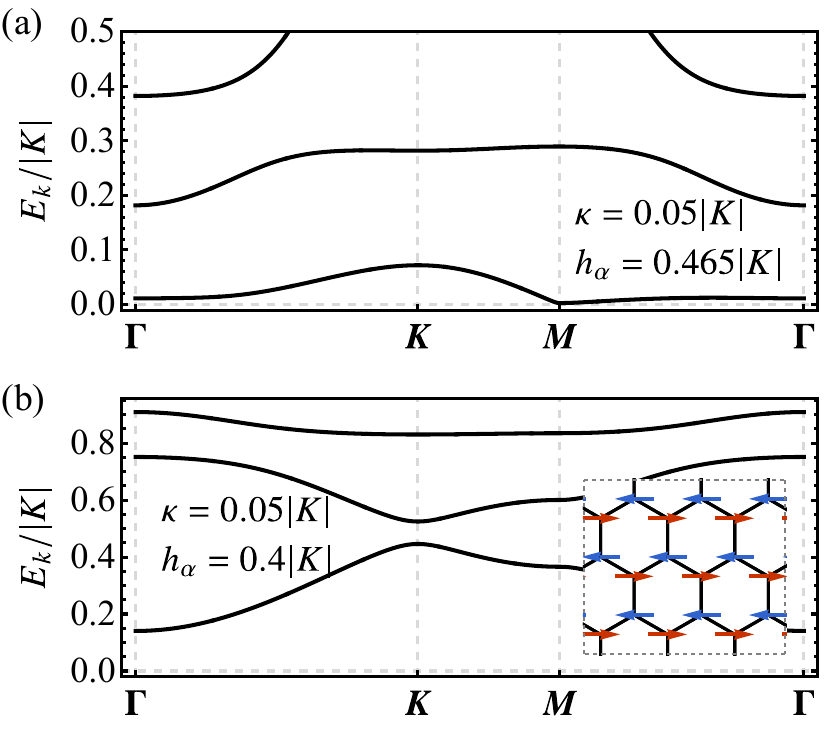}
\caption{
Fermion and boson bands for the AFM model.
(a) Fermion bands in the narrow intermediate $C = -2$ regime, note that the lowest
band has a tiny gap and is very flat, thus contributing a large low-energy density of states.
(b) Bosonic vison pair bands. The inset illustrates the pattern of possible in-plane N\'eel order
when the soft $\bGamma$ mode condense.
}\label{fig:AFM-pair-bands}
\end{figure}

As $h$ increases, the fermion bands undergo topological phase transitions accompanied by gap closings.
In the AFM model, the pair hopping $t^\chi_{\alpha,\beta}$ is relatively small.
At small $h$, the Chern number $C = 1$ and the band bottom is initially at $\bK$ but gradually shifts to the $\bM$ point as $h$ increases,
first closing at $h^\mathrm{AFM}_{\chi 1} \approx 0.4-0.6$, depending on $\kappa$.
This results in a transition into a state with Chern number $C = -2$ (see the bands in \Fref{fig:AFM-pair-bands}(a)).
Interestingly, this state exists over a very narrow window, and at a subsequent $h^\mathrm{AFM}_{\chi 2}$,
the gap closes at the $\bGamma$ point, driving a transition again into a state with $C = 1$.
Finally, at a higher $h^\mathrm{AFM}_{\chi 3}$, the band gap closes again at the $\bGamma$ point, rendering the band topology 
trivial ($C = 0$) thereafter. A plot of $C$ at different $(\kappa, h)$ values is shown in \Fref{fig:phase-diagram}(b).
Notably, the states with $C = 1$ at small and intermediate $h$ have \emph{opposite} topological parity indices $\zeta_{\bk}$ 
at the four high-symmetry momenta~\cite{Kou2009-dm}:
the low-$h$ AFM-KSL phase has $(\zeta_{\bGamma}, \zeta_{\bM_1}, \zeta_{\bM_2}, \zeta_{\bM_3}) = (0,1,1,1)$,
whereas the intermediate state exhibits $(\zeta_{\bGamma}, \zeta_{\bM_1}, \zeta_{\bM_2}, \zeta_{\bM_3}) = (1,0,0,0)$.
Interestingly, the topological indices of the latter match those of the FM-KSL 
(see \Appref{sec:feremion_parity})~\cite{Chen2023-vo}.
The physical significance of the intermediate $C = 1$ state is currently unclear to us, since it appears beyond the closing of the vison 
gap \footnote{However, it is noteworthy that it is reminiscent of the $C = 1$ parton state state from Ref.\cite{Jiang2020-mo},
which was argued to lack topological order upon performing exact Gutzwiller projection.}. In the FM model, due to the stronger pair hopping,
we have found that at a small $h^\mathrm{FM}_{\chi}$, the fermion band gap closes at the $\bGamma$ point,
resulting in $C \rightarrow 0$ and the loss of band topology, which is consistent with a gauge confinement transition into the SP phase.
It should be noted that, due to the perturbative nature of our calculations, the obtained fermion Chern numbers are 
most reliable in the low-to-intermediate field regime.

\section{Bosons}
The bosonic vison pairs can be constructed as:
\begin{equation} \label{eq:boson-pair-def}
    d_{\br,\alpha}^\dagger | \Omega \rangle = \chi_{\br,\alpha}^\dagger \tilde{\alpha}_1^\dagger |0; \Psi^c_0(\br,\alpha) \rangle.
\end{equation}
Here $\tilde{\alpha}_j^\dagger$ creates the $j$-th Bogoliubov quasiparticle (Bogolon) of the $c$-Majorana $H^c(\br,\alpha)$,
and $\tilde{\alpha}_1^\dagger |0; \Psi^c_0(\br,\alpha) \rangle$ is the lowest energy state that has an odd parity of $c$-Majoranas,
so $d_{\br,\alpha}^\dagger |\Omega \rangle$ has an \emph{even} total fermion parity,
i.e., the same as the ground state $| \Omega \rangle$~\cite{Knolle2014-wl,Knolle2015-lx,Zhang2021-xi}.
Therefore $d^\dagger_{\br,\alpha}$ adds a local boson excitation to the absolute vacuum,
i.e., a magnon-like quasiparticle made from a bound state of the fermion vison-pair $\chi^\dagger_{\br,\alpha}$ and a matter $c$-fermion. 
To leading order, the Zeeman field induces bosonic vison-pair hopping and creation/annihilation
(see \Appref{sec:bosonic-pair} for more details).
Therefore, the
effective boson Hamiltonian reads:
\begin{align} \label{eq:H_boson}
    H_\mathrm{B} = & \sum_{\alpha \neq \beta} t^d_{\alpha,\beta} \left( \sum_{\br \in A} d_{\br,\alpha}^\dagger d_{\br,\beta}
    + \sum_{\br \in B} d_{\br-\bdelta_{\alpha},\alpha}^\dagger d_{\br-\bdelta_\beta,\beta} \right) \nonumber \\
    & + \sum_{\br,\alpha} \lambda_d (d_{\br,\alpha}^\dagger + d_{\br,\alpha}) + \Delta_d \, d_{\br,\alpha}^\dagger d_{\br,\alpha},
\end{align}
with $t^d_{\alpha, \beta} = (t^d_{\beta, \alpha})^*$ and $\lambda_d \in \mathbb{R}$.
Similar to fermionic pairs, we have found that the bosonic pair hopping $t^d_{\alpha, \beta}$ in the FM case is significantly stronger than
that of the AFM case.
Interestingly, we have found that $\lambda_d$ is finite in the FM case and vanishes in the AFM case.
$\lambda_d$ is defined as:
\begin{align} \label{eq:lambda_d}
 \lambda_d \equiv &\langle \Omega | H_h d_{\br,\alpha}^\dagger | \Omega \rangle \nonumber \\
 = & (-h_\alpha) \langle \Omega | ( \sigma_{\br}^\alpha + \sigma_{\br+\bdelta_\alpha}^\alpha ) d_{\br, \alpha}^\dagger | \Omega \rangle.
\end{align}
Therefore the vanishing $\lambda_d$ in the AFM case is an indication that its lowest energy boson carries different symmetry representations
than the $[111]$ Zeeman field operator.
In fact, we have found that the boson pair in the AFM model, $d_{\br,\alpha}^\dagger |\Omega \rangle$, is \emph{odd} under spatial inversion 
about bond $(\br,\alpha)$'s center, while the Zeeman field operator is even,
leading to a cancellation of the two terms in \Eqref{eq:lambda_d}.
On the other hand, these two terms are identical and add up in the FM case, implying that the boson pair in this case is even under inversion.
%
%
In order to fully characterize the potential patterns of spontaneous symmetry breaking resulting from this boson condensation, 
we consider its transformations under other symmetries.
In both cases, we found that the critical boson modes have momentum $\bk = 0$,
therefore we do not expect spontaneous breaking of lattice translations.
In the FM case, the critical mode is an equal superposition of the three types of pairs:
$\beta_{\bGamma,1}^\dagger = ( d_{\bGamma,x}^\dagger + d_{\bGamma,y}^\dagger + d_{\bGamma,z}^\dagger )/\sqrt{3}$,
and therefore transforms trivially under $C_3$ rotations.
Thus the soft boson mode in the FM case carries trivial quantum numbers and transforms like the $[111]$ Zeeman operator, and there is no associated spontaneous symmetry breaking but its softening is simply an indication of the \emph{continuous} development of spin polarization along the $[111]$ direction.
On the other hand, for the AFM model the critical mode has an orbital angular momentum $1\, (\mathrm{mod}\, 3)$:
$\beta_{\bGamma,1}^\dagger = ( d_{\bGamma,x}^\dagger + e^{i 2\pi/3} d_{\bGamma,y}^\dagger + e^{i 4\pi/3} d_{\bGamma,z}^\dagger )/\sqrt{3}$.
In fact, the soft boson mode in the AFM case transforms  as a N\'{e}el order parameter with spin moments orthogonal to the $[111]$ direction, and thus we expect its condensation to drive  true spontaneous symmetry breaking with spin moments given by
$\langle \bm{\sigma}_{\br} \rangle \propto ( \cos(\theta), \cos(\theta+2\pi/3), \cos(\theta+4\pi/3) ) \perp (1,1,1)$ for $\br \in A$,
and opposite orientation for $r\in B$, as depicted in \Fref{fig:AFM-pair-bands}(b) (see also \Appref{sec:bosonic-pair}).
Therefore our analysis of soft vison-boson pairs suggests a direct phase transition from the KSL to SP in the FM model,
and towards an AFM phase with in-plane magnetic canting in the AFM model.
Moreover, since these bosons move on a Kagome lattice, their band topology is non-trivial.
The lowest-energy boson band in the FM (AFM) model has Chern number $1$ ($-1$),
reminiscent of the Chern number obtained for magnons at high fields from spin-wave theory analysis~\cite{McClarty2018-po,Joshi2018-os,Chern2021-gc}.

It is worth noting that the distinct behavior of bosonic vison pairs also manifests in the dynamical signatures of the Kitaev model.
For instance, in the FM Kitaev model, the expectation value
$\langle \Omega | \sigma_{\bm{q}=0}^\alpha d_{\bm{r},\alpha}^\dagger | \Omega \rangle = 
\langle \Omega | ( \sigma_{\bm{r}}^\alpha + \sigma_{\bm{r}+\bm{\delta}_\alpha}^\alpha )  d_{\bm{r},\alpha}^\dagger | \Omega \rangle$
is finite, whereas it vanishes in the AFM case.
This difference directly correlates with the presence (in FM) or absence (in AFM) of a coherent peak near the vison-pair excitation energy 
in the dynamical structure factor $S(\omega, \bm{q}=0)$~\cite{Knolle2015-lx}. Additionally, the stronger spectral weight at the 
$\bm{\Gamma}$ point in the FM Kitaev model suggests a greater tendency to transition into the SP state under a uniform Zeeman field.

\section{Discussion}
By examining different quasiparticles of the Kitaev model (visons, fermions, and bosons), we found that they become gapless at critical values of 
the Zeeman field that are qualitatively consistent with numerical findings~\cite{Gohlke2018-kg,Zhu2018,Hickey2019,Patel2019,Wang2024-ku,Zhu2024-ti},
as shown in \Fref{fig:phase-diagram}.
In the FM model, both the visons and the fermions become gapless at nearly the same value of the Zeeman field.
The closing of the fermion gap drives a transition into a state with zero Chern number, rendering the visons into ordinary bosons capable of condensing.
This suggests that the KSL-SP transition at small $h^\mathrm{FM}_c$
possibly occurs via a non-trivial quantum critical point as argued in Ref.~\cite{Zou2020-ni}.
Remarkably, for the AFM model we have found that all three kinds of quasi-particles
become gapless at very similar critical values of the Zeeman field (see \Fref{fig:phase-diagram}(b)).
Previous studies that have proposed theories for these transitions by focusing only on the fermions are unlikely to capture the nature of the IP,
as they presume that single visons are gapped and only the fermions are becoming gapless.
Furthermore, in the AFM Kitaev model, a soft boson mode that also becomes gapless at similar critical fields and
carries the quantum numbers of an in-plane N\'{e}el order (as illustrated in the inset of \Fref{fig:AFM-pair-bands}(b)).
Interestingly, similar AFM correlations were observed in a recent machine learning study~\cite{Zhang2024-ec}.
This raises the possibility that the IP may exhibit in-plane AFM magnetization, consistent with magnetic canting,
given the presence of a dominant out-of-plane moment~\cite{Wang2024-ku}.
While explicit numerical evidence for in-plane AFM order is still lacking, 
indirect indications exist, such as the softening of a coherent magnon mode at the $\Gamma'$ point as the system transitions from 
SP to IP~\cite{Gohlke2018-kg,Hickey2019}. More systematic numerical studies would be valuable in clarifying this issue.

\acknowledgments
We are thankful to Penghao Zhu, Kang Wang, Nandini Trivedi, and Jiucai Wang for valuable discussions, and we extend sepcial thanks to
Shangshun Zhang, G\'{a}bor H\'{a}lasz, and Christian Batista for several patient and dedicated discussions on details of 
their works.
I.S.V. acknowledges support from the Deutsche Forschungsgemeinschaft (DFG) under research Grant No.~542614019.
C.C. acknowledges support from the National Natural Science Foundation of China under Grants No.~12404175 and No.~12247101,
the Fundamental Research Funds for the Central Universities (Grant No.~lzujbky-2024-jdzx06), 
and the Natural Science Foundation of Gansu Province (No.~22JR5RA389).
We thank Beijing Paratera Co., Ltd. for providing HPC resources that contributed to the numerical results reported in this paper.

\appendix

\section{Single visons}

\subsection{Single-vison bands}
\begin{figure}
\centering
\includegraphics[width=0.49 \textwidth]{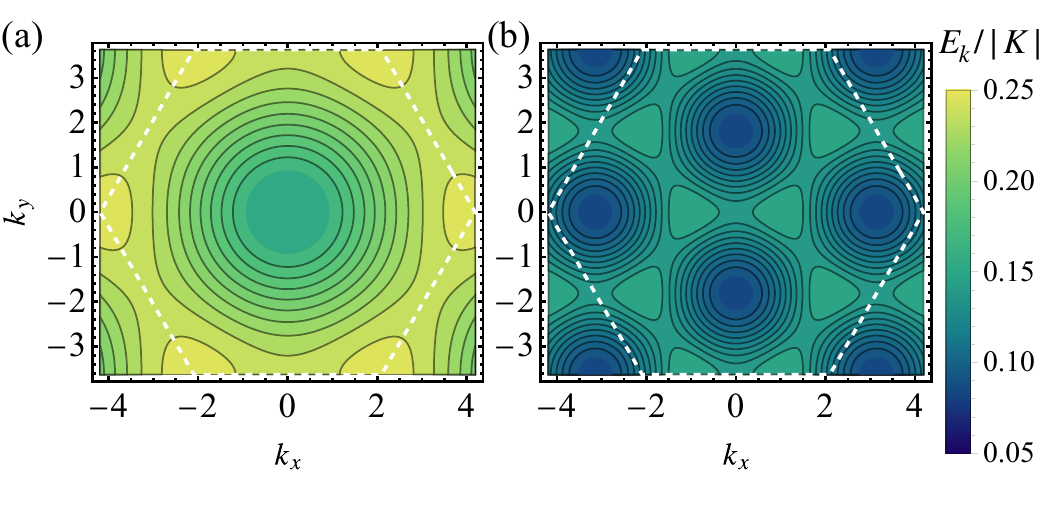}
\caption{
Single visons' energy bands.
(a) Vison band in the FM Kitaev model ($\kappa = 0.05|K|, h_\alpha = 0.01 |K|$).
Here the band minimum is at the $\bGamma$ point. The white dashed hexagon indicates
the Brillouin zone of the honeycomb lattice.
(b) (Lower) Vison band in the AFM Kitaev model ($\kappa = 0.05|K|, h_\alpha = 0.3 |K|$).
Here the band minimum is at $\bM/2$.
}\label{fig:vison-bands}
\end{figure}

As discussed in Ref.~\cite{Chen2023-vo}, with FM Kitaev coupling, the Zeeman induced vison hopping
preserves the translational symmetry of the honeycomb lattice.
However, in the AFM Kitaev model, visons experience a uniform $\pi$-flux within each honeycomb unit cell,
meaning that the trasnlational symmetry is only implemented projectively.
In this case, we find that there are two vison bands with non-trivail topology (Chern number $C = \pm 1$),
and the minimum of the lower band is located at $\bM/2$.
\Fref{fig:vison-bands} contains plots of vison bands at selected $(\kappa,h)$ values in both cases.

\begin{figure}
\centering
\includegraphics[width=0.45 \textwidth]{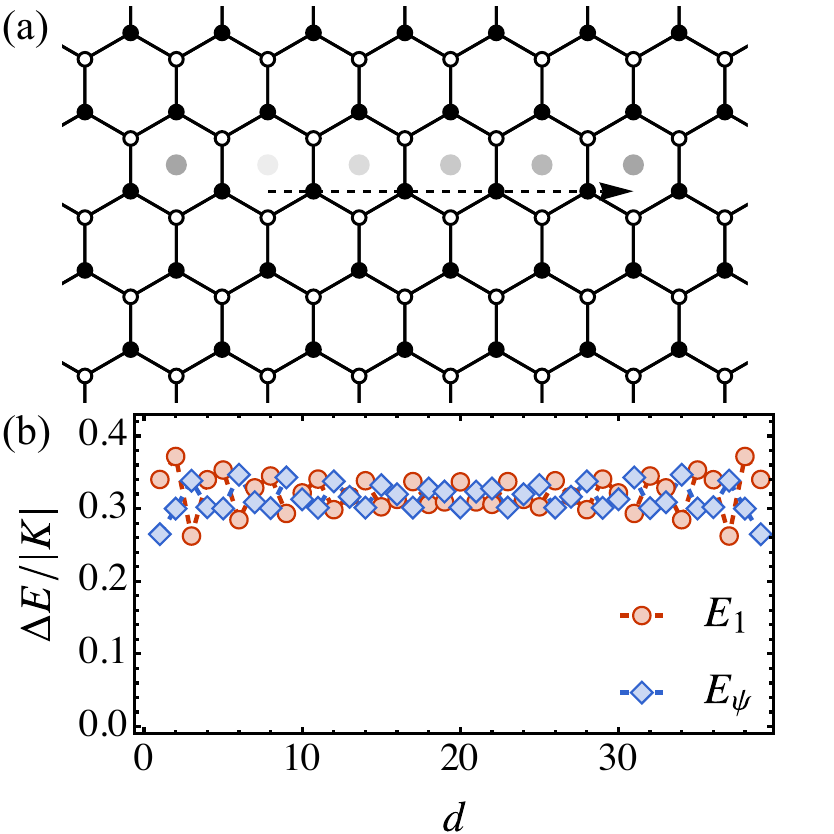}
\caption{
(a) Schematic of two visons seprating from each other.
(b) The excitation energy of two types of $2$-vison states as a function of their separation:
$E_1$ represents for the bosonic $2$-vison state, and $E_\psi$ stands for the fermionic $2$-vison states.
The fermionic vison pair state ($E_\psi$ at $d = 1$) has the lowest excitation energy.
Note that the data is generated on a $40 \times 40$ torus, so their maximum separation is at $d = 20$.
Here $\kappa = 0.01 |K|$.
}\label{fig:2-vison}
\end{figure}

\subsection{Energy of two-vison states} \label{sec:E_two-vison}
A pair of visons experience a weak attractive interaction between them.
\Fref{fig:2-vison} shows the excitation energy of the lowest energy bosonic ($E_1$) and fermionic ($E_\psi$) $2$-vison
states as function of their separation.
Here the data is generated for a 40 × 40 torus, so the maximum separation between the two visons
(with maximum $\Delta E$) is at $d = 20$.
The nearest-neighbor fermionic $2$-vison states, i.e., the fermionic vison pair state defined in the main text,
have the lowest energy, justifying their inclusion in the low-energy effective model.

\section{Fermionic vison pairs and itinerant Majoranas}\label{sec:compare}
%
\begin{figure*}
\centering
\includegraphics[width=0.7 \textwidth]{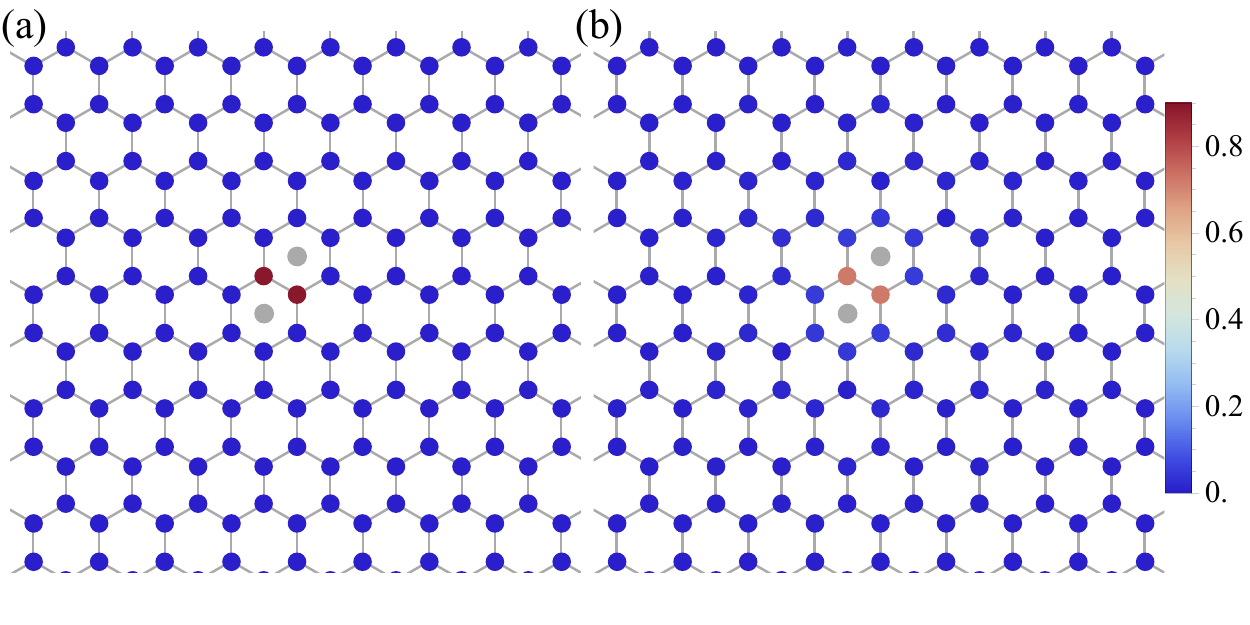}
\caption{
Real-space couplings (amplitude) between $\tilde{\chi}_\mu$ and $c$ Majoranas, obtained
(a) in this work and (b) from Ref.~\cite{Zhang2022-tc},
using the example of a $\tilde{\chi}_{\br,y}$ vison pair (the gray dots stands for the visons).
The $p_{\bR}/h_\alpha$ found in this study is highly localized, while that defined in Ref.~\cite{Zhang2022-tc}
is much more extended. 
For example, the couplings to the Majoranas separating the vison pair are on the order of 
$10^{-10}$ in our case, whereas the couplings to the Majoranas located two lattice constants away 
from the vison pair remain approximately $0.01$.
The data presented here is generated at $\kappa = 0$.
}\label{fig:p_R}
\end{figure*}

\begin{figure*}
\centering
\includegraphics[width=0.75 \textwidth]{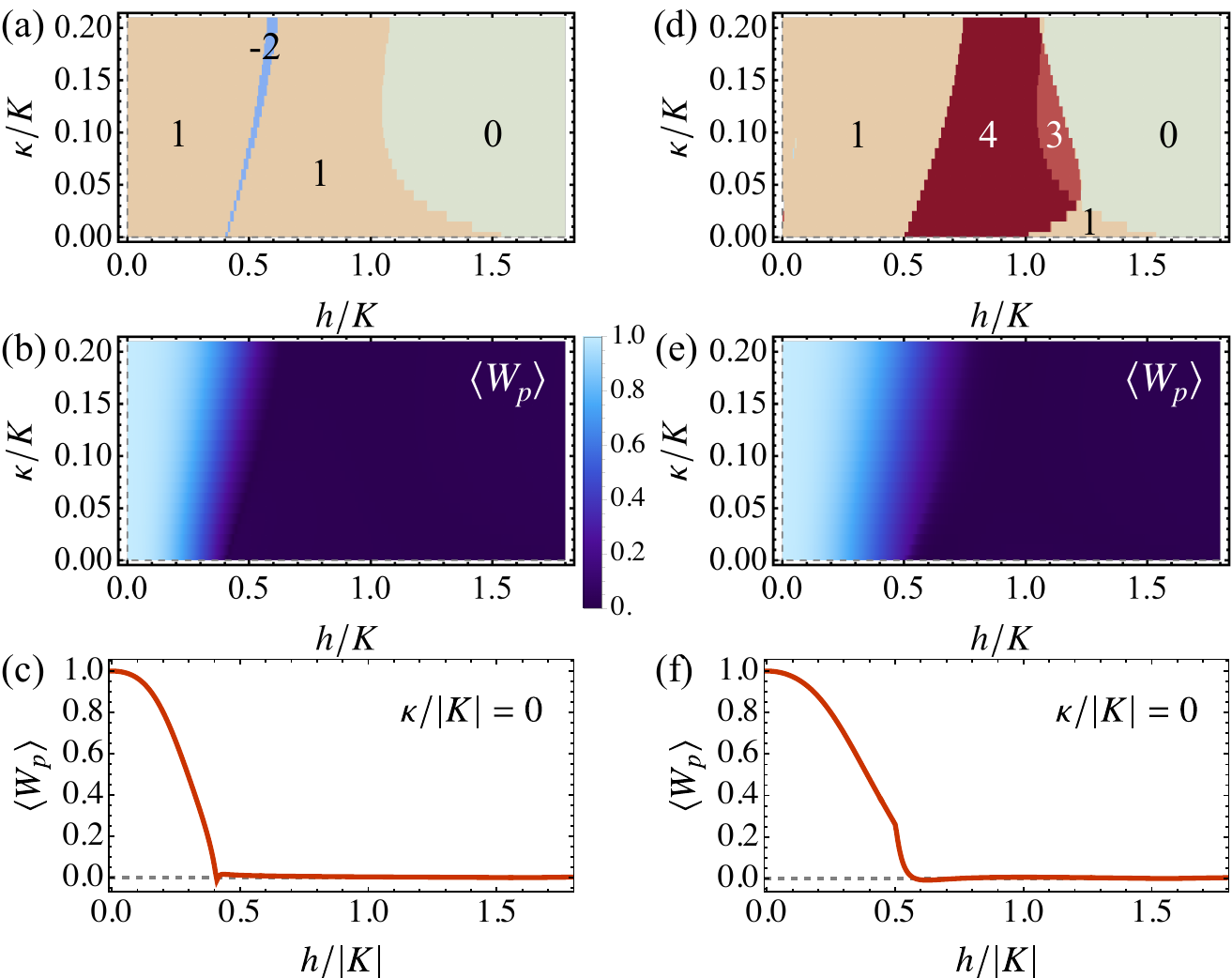}
\caption{
Comparison of the fermion sector results between this work and Ref.~\cite{Zhang2022-tc}
(reproduced in this study).
(a)-(c) Results obtained from \Eqref{eq:chi-c-hyb}:
The fermion Chern number, vison parity $\langle W_p \rangle$ as a function of $(\kappa,h)$,
and $\langle W_p \rangle$ v.s. $h$ at $\kappa = 0$.
Note the $\langle W_p \rangle$ shown here only takes into account the contribution from fermion vison pairs,
as was done in Ref.~\cite{Zhang2022-tc}.
(d)-(f) The same type of quantities, reproduced using the method from Ref.~\cite{Zhang2022-tc}.
Note that Ref.~\cite{Zhang2022-tc} considered only the $\kappa = 0$ case.
}\label{fig:compare}
\end{figure*}
\subsection{Determining the coupling constants}
In this section, we discuss the determination of coupling constants $t^\chi_{\alpha, \beta}$ and $p_{\bR}$
in \Eqref{eq:H_fermion} of the main text. Moreover, we will compare the $p_{\bR}$ defined in our convention and
that from Ref.~\cite{Zhang2022-tc}.

The fermionic vison pair hopping is defined as the matrix transition amplitude for the Zeeman perturbation,
$H_h$, between the two eigenstates of the $K-\kappa$ model with initial and final pair configurations respectively,
which explicitly is given by:
\begin{equation} \label{eq:t^chi-def}
    t^\chi_{\alpha, \beta} \equiv \langle\Omega| \tilde{\chi}_{\br,\alpha} H_h \tilde{\chi}_{\br, \beta}^\dagger|\Omega\rangle.
\end{equation}
Here we illustrate the calculation by taking the example of $t^\chi_{x,y}$.
Using Kitaev's Majorana representation, the intial and final states are given by:
\begin{subequations} \label{eq:f-vpair-states}
\begin{align}
    \tilde{\chi}_{\br,x}^\dagger |\Omega \rangle = \chi_{\br,x}^\dagger |0; \Psi^c_0(\br,x) \rangle, \\
    \tilde{\chi}_{\br,y}^\dagger |\Omega \rangle = \chi_{\br,y}^\dagger |0; \Psi^c_0(\br,y) \rangle.
\end{align}
\end{subequations}
Here $|0;\Psi^c_0(\br,\alpha) \rangle$ is the ground state of the $c$-Majorana BdG Hamiltonian $H^c(\br,\alpha)$,
which contains a pair of neighboring visons obtained by flipping the sign of $u_{\br,\br+\bdelta_\alpha}$.
Note the state $|0; \Psi^c_0(\br, \alpha) \rangle$ is parity \emph{even}, i.e.,
same as that of the ground state without any fluxes~\cite{Knolle2014-wl,Zhang2021-xi}.
In \Eqref{eq:t^chi-def}, only two terms contribute: $-h_z \sigma_{\br}^z \leftrightarrow (-h_z) (-i b_{\br,x} b_{\br,y} )$
and $-h_z \sigma_{\br+\bdelta_z}^z \leftrightarrow (-h_z) (-i b_{\br}^x b_{\br}^y) i c_{\br} c_{\br+\bdelta_z}
(-i b_{\br}^z b_{\br+\bdelta_z}^z)$. Therefore $t^\chi_{x,y}$ reads:
\begin{equation} \label{eq:t^chi_z}
    t^\chi_{x,y} = (i h_z) \langle 0; \Psi^c_0(\br,x) | (1+ic_{\br} c_{\br+\bdelta_z}) | 0; \Psi^c_0(\br,y) \rangle,
\end{equation}
and the overlap can be calculated by standard methods~\cite{Robledo2009-ze,Knolle2014-wl,Zhang2022-tc,Chen2023-vo}.

In order to obtain the $p_{\bR}$, which quantifies the mixing of matter majoranas and fermionic vison-pairs (see Eq.~(4) of main text),
it is useful to first rewrite (for each $\mu$)
the $\tilde{\chi}_\mu$-$c$ hybridization part of Eq.~(4)
as:
\begin{align}a\label{eq:chi-c-hyb}
    H_{\tilde{\chi}_\mu-c} = \sum_{\br} \sum_{j} ( \lambda^{\br,\mu}_j \, \alpha_j^\dagger \tilde{\chi}_{\br,\mu}
    + \eta^{\br,\mu}_j \, \tilde{\chi}_{\br,\mu} \tilde{\alpha}_j + h.c. )
\end{align}
Here, $\alpha_j$ is the $j$-th Bogolon destruction operator made from a linear combination of c-matter Majoranas associated with
$c$-BdG Hamiltonian with no vison, $H^c(0)$,
and $\tilde{\alpha}_j$ is the destruction operator of the $j$-th Bogolon of the $c$-BdG Hamiltonian in gauge sector of a $(\br,\mu)$
vison pair, $H^c(\br,\mu)$.
The $\lambda^{\br,\mu}_j$ quantifies the tunneling between a fermion vison pair and a $c$-Bogolon of the zero-flux
sector and it is explicitly defined by the following transition amplitude computed from the Zeeman perturbation ($H_h$):
\begin{align} \label{eq:lambda-def}
    \lambda^{\br,\mu}_j = & \langle \Omega | \alpha_j H_h \tilde{\chi}_{\br,\mu}^\dagger | \Omega \rangle \nonumber \\
    = & \langle 0; \Psi^c_0(0) | \alpha_j H_h \chi_{\br,\mu}^\dagger | 0; \Psi^c_0(\br,\mu) \rangle.
\end{align}
The $\eta^{\br,\mu}_j$ quantifies the pair creation of a fermion vison pair and a $c$-Bogolon within the same gauge sector
and it is defined as:
\begin{align}
    \eta^{\br,\mu}_j = & \langle \Omega | H_h \tilde{\alpha}_j^\dagger \tilde{\chi}_{\br,\mu}^\dagger | \Omega \rangle \nonumber \\
    = & \langle 0; \Psi^c_0(0) | H_h \tilde{\alpha}_{j}^\dagger \chi_{\br,\mu}^\dagger |0; \Psi^c_{0}(\br,\mu) \rangle.
\end{align}
After obtaining the $\lambda^{\br,\mu}$ and $\eta^{\br,\mu}$ numerically, one can rewrite the $\alpha$- and $\tilde{\alpha}$-Bogolons
in terms of local $c_r$ Majoranas and obtain the $p_{\bR}$ parameters in Eq.~(5).
We have found that the $p_{\bR}$ obtained this way is a highly \emph{local} function of $\bR$,
see \Fref{fig:p_R}(a) for a plot of the real-space couplings (divided by $-h$)
between a $\tilde{\chi}_{\br,z}$ pair and the $c$-Majoranas.

In Ref.~\cite{Zhang2022-tc}, the $\tilde{\chi}-c$ hybridization is defined slightly differently.
In that case, the tunneling between $\tilde{\chi}$ and $c$-Bogolons of the $0$-vison sector ($\alpha_j^\dagger \tilde{\chi}_{\br,\mu}$)
is the same as in \Eqref{eq:chi-c-hyb}.
Note that there, $\psi_{\bk}$ is used to represent the Bogolons of the $0$-vison sector, owing to the translational symmetry of $H^c(0)$.
On the other hand, the pairing part in Ref.~\cite{Zhang2022-tc} is defined as $\zeta^{\br,\mu}_j \tilde{\chi}_{\br,\mu} \alpha_j$,
also using the Bogolons of the $0$-flux sector, and $\zeta^{\br, \mu}_j$ is taken to be fixed by
(cf. Eqs.~(S2) and (S3) in the Supplimentary Information of Ref.~\cite{Zhang2022-tc}):
\begin{align} \label{eq:zeta-def}
    \zeta^{\br,\mu}_j = \langle \Omega | H_h \alpha_j^\dagger \tilde{\chi}_{\br,\mu}^\dagger | \Omega \rangle.
\end{align}
However, $\alpha_j^\dagger \tilde{\chi}_{\br,\mu}^\dagger | \Omega \rangle$ (denoted as $\psi_{-\bk}^\dagger \tilde{\chi}_{\br,\mu}^\dagger | \Omega \rangle$
in Ref.~\cite{Zhang2022-tc}) is \emph{not} an eigenstate of the model, and states $\alpha_j^\dagger \tilde{\chi}_{\br,\mu}^\dagger |\Omega \rangle$
with different $j$ are not orthogonal. Thus, the right-hand side of \Eqref{eq:zeta-def} cannot strictly speaking be interpreted as
a transition amplitude induced by the Zeeman perturbation between the eigenstates of the exactly solvable model in the absence of Zeeman field.
Therefore, we believe that \Eqref{eq:chi-c-hyb}, where the coefficients are determined by matrix elements of the 
Zeeman perturbation between an \emph{eigenbasis} of the Kitaev model, involving:
$\tilde{\chi}_{\br,\mu}^\dagger | \Omega \rangle$, $\alpha_j^\dagger | \Omega \rangle$,
and $\tilde{\alpha}_j^\dagger \tilde{\chi}_{\br,\mu}^\dagger | \Omega \rangle$,
provides a more satistactory description of the coupling between $\tilde{\chi}_\mu$-fermions and itinerant $c$-Majoranas.
Nonetheless, we have reproduced the calculations from Ref.~\cite{Zhang2022-tc} and presented relevant results here for comparison.
See \Fref{fig:p_R}(b) for the reproduced $p_{\bR}$ and \Fref{fig:compare} for a comparison of the fermion
Chern numbers and vison parity $\langle W_p \rangle$ at different $(\kappa,h)$ values between the two studies.
Note that Ref.~\cite{Zhang2022-tc} only considered the case of $\kappa = 0$, but this does not affect substantially 
the calculation of fermion spectrum as sensitively as it affects the single vison properties.

\begin{figure}
\centering
\includegraphics[width=0.45 \textwidth]{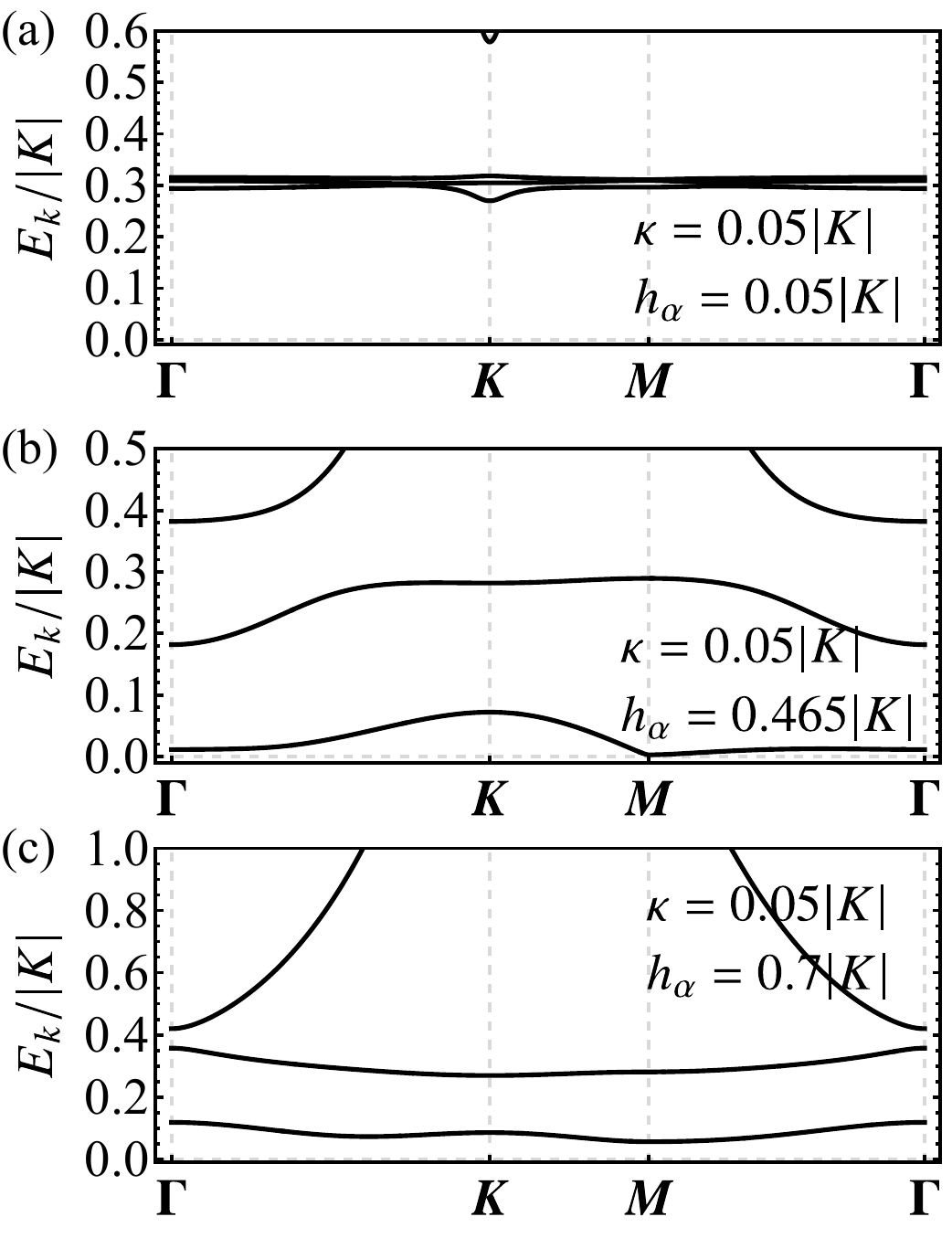}
\caption{
Fermion bands (only showing those with $E>0$) for AFM Kitaev model at selected points with
$\kappa = 0.05 |K|$ (represented by the blue stars in \Fref{fig:full-phase-diagram}(b)).
(a) Bands in the KSL phase ($h_\alpha = 0.05 |K|$), the Chern number $C = 1$.
(b) Bands in the tiny intermediate $C = -2$ regime ($h_\alpha = 0.465 |K|$), note that the lowest
band has a tiny gap and is rather flat, thus contributing a large low-energy density of states.
(c) Bands in the intermediate $C = 1$ state ($h_\alpha = 0.7 |K|$). Note the lowest band
is rather flat in this regime.
}\label{fig:f-bands}
\end{figure}

The fermion bands at selected $(h,\kappa)$ points (highlighted by the blue stars in \Fref{fig:full-phase-diagram}(b)) 
computed within our approach are shown in \Fref{fig:f-bands}.
As $h$ increases, the band gap initially forms at the $\bK$ point with Chern number $C = 1$.
The band bottom gradually shifts to the $\bM$ point as $h$ increases,
first closing at $h^\mathrm{AFM}_{\chi 1} \approx 0.4-0.6$, depending on $\kappa$.
This results in a transition into a state with Chern number $C = -2$.
Interestingly, this state exists over a very narrow window, and at a nearby $h^\mathrm{AFM}_{\chi 2}$,
the gap closes at the $\bGamma$ point, driving a transition again into a state with $C = 1$.
Finally, at a higher $h^\mathrm{AFM}_{\chi 3}$, the band gap closes again at the $\bGamma$ point, rendering the band topology 
trivial ($C = 0$) thereafter.

\subsection{Fermion pairty indices} \label{sec:feremion_parity}
The parity indices of matter fermions at the four high-symmetry $k$-points, $\{ \zeta_{\bGamma}, \zeta_{\bM_1}, \zeta_{\bM_2}, \zeta_{\bM_3} \}$,
provide valuable information for classifying QSLs~\cite{Kou2009-dm,Kou2010-de,Rao2021-xa}.
As discussed in Ref.~\cite{Chen2023-vo}, the matter fermion's parity indices for the FM and AFM KSL
are $\{ 1,0,0,0 \}$ and $\{ 0,1,1,1 \}$, respectively.
In the Kitaev model under a Zeeman field, it was found that whenever the fermion bands undergo a gap closing at certain 
high-symmetry $k$-points, the parity indices at those $k$-points change by $1$ ($\mathrm{mod} \ 2$) accordingly.

For the FM Kitaev model, at $h_{\chi}^\mathrm{FM}$, the fermion gap closes at $\bGamma$ point, $\zeta_{\bGamma} \rightarrow 0$,
making the parity indices trivial ($\{ 0,0,0,0 \}$) thereafter.
In contrast, for the AFM Kitaev mode, at $h_{\chi 1}^\mathrm{AFM}$, gap closes at all three $\bM$ points, changing the parity indices
to $\{ 0,0,0,0 \}$. Shortly after, at $h_{\chi 2}^\mathrm{AFM}$, the gap closes at $\bGamma$ point, shifting the indices to
$\{ 1,0,0,0 \}$, identical to that of the FM KSL. Finally, at $h_{\chi 3}^\mathrm{AFM}$, the gap closes at $\bGamma$ point again
and the parity indices become trivial ($\{ 0,0,0,0 \}$).

\section{Bosonic vison pairs} \label{sec:bosonic-pair}
\subsection{Band energy and topology}
Similar to the fermionic case, the hopping of bosonic vison pairs $t^d_{\alpha,\beta}$ is defined as:
\begin{align}
    t^d_{\alpha, \beta} \equiv \langle \Omega | d_{\br,\alpha} H_h d_{\br,\beta}^\dagger | \Omega \rangle.
\end{align}
Since a single-bosonic-pair state is defined as:
$d_{\br,\mu}^\dagger | \Omega \rangle \equiv \chi_{\br,\mu}^\dagger \tilde{\alpha}_1^\dagger(\br,\mu) | 0; \Psi^c_0(\br,\mu) \rangle$,
with $\tilde{\alpha}_1^\dagger(\br,\mu)$ being the creation operator for the lowest energy Bogolon of $c$-BdG Hamiltonian $H^c(\br,\mu)$.
It can be shown that the boson pair hopping reduced to:
\begin{align} \label{eq:expr-t^d}
    t^d_{\alpha, \beta} = & \varepsilon_{\alpha \beta \gamma} (ih_\gamma) \langle 0; \Psi^c_0(\br,\alpha) |
    \tilde{\alpha}_1(\br,\alpha) (1 + i c_{\br} c_{\br+\bdelta_\gamma}) \nonumber \\
    & \tilde{\alpha}_1^\dagger(\br,\beta) | 0; \Psi^c_0(\br, \beta) \rangle.
\end{align}
Here $\varepsilon_{\alpha \beta \gamma}$ is the Levi-Civita symbol.
The single-boson creation/annihilation amplitude $\lambda_d$ is defined as:
\begin{align}\label{eq:lambda-def}
    \lambda_d = & \langle \Omega | H_h d_{\br,\mu}^\dagger | \Omega \rangle \nonumber \\
    = & - h_\mu \langle 0; \Psi^c_0(0) | ( i c_{\br} + c_{\br+ \bdelta_\mu} ) \tilde{\alpha}_1^\dagger(\br,\mu)
    | 0; \Psi^c_0(\br,\mu) \rangle.
\end{align}
We have found that for FM Kitaev model $\lambda_d \neq 0$, whereas for AFM Kitaev model $\lambda_d = 0$.
This is related to the fact that under the spatial inversion transformation ($\mathcal{I}$) around the center of bond $(\br,\mu)$,
$\tilde{\alpha}_1^\dagger(\br,\mu) \rightarrow -i \tilde{\alpha}_1^\dagger(\br,\mu)$ in the FM Kitaev model,
and $\tilde{\alpha}_1^\dagger(\br,\mu) \rightarrow i \tilde{\alpha}_1^\dagger(\br,\mu)$ in the AFM Kitaev model.
Combined with the fact that under $\mathcal{I}$, $\chi_{\br,\mu} \rightarrow i \chi_{\br,\mu}$,
$| \Psi^c_0 (\br,\mu) \rangle \rightarrow | \Psi^c_0(\br,\mu) \rangle$.
It can be seen that the bosonic vison pair state $d_{\br,\alpha}^\dagger | \Omega \rangle$ is even (odd)
under $\mathcal{I}$ in the FM (AFM) Kitaev model.
This immediately indicates that:
$\langle \Omega | \sigma_{\br}^\mu d_{\br,\mu}^\dagger | \Omega \rangle =
\langle \Omega | \sigma_{\br+\bdelta_\mu}^\mu d_{\br,\mu}^\dagger | \Omega \rangle$
for the FM Kitaev model, and 
$\langle \Omega | \sigma_{\br}^\mu d_{\br,\mu}^\dagger | \Omega \rangle = 
-\langle \Omega | \sigma_{\br+\bdelta_\mu}^\mu d_{\br,\mu}^\dagger | \Omega \rangle$
for the AFM Kitaev model.

Note that under $\mathcal{I}$:
\begin{align} \label{eq:inv-trans}
    & c_{\br} \rightarrow c_{\mathcal{I}(\br)}, \ c_{\mathcal{I}(\br)} \rightarrow -c_{\br}, \nonumber \\
    & b_{\br}^\mu \rightarrow b_{\mathcal{I}(\br)}^\mu, \ b_{\mathcal{I}(\br)}^\mu \rightarrow -b_{\br}^\mu,
\end{align}
with $\br \in A$.

\begin{figure}
\centering
\includegraphics[width=0.49 \textwidth]{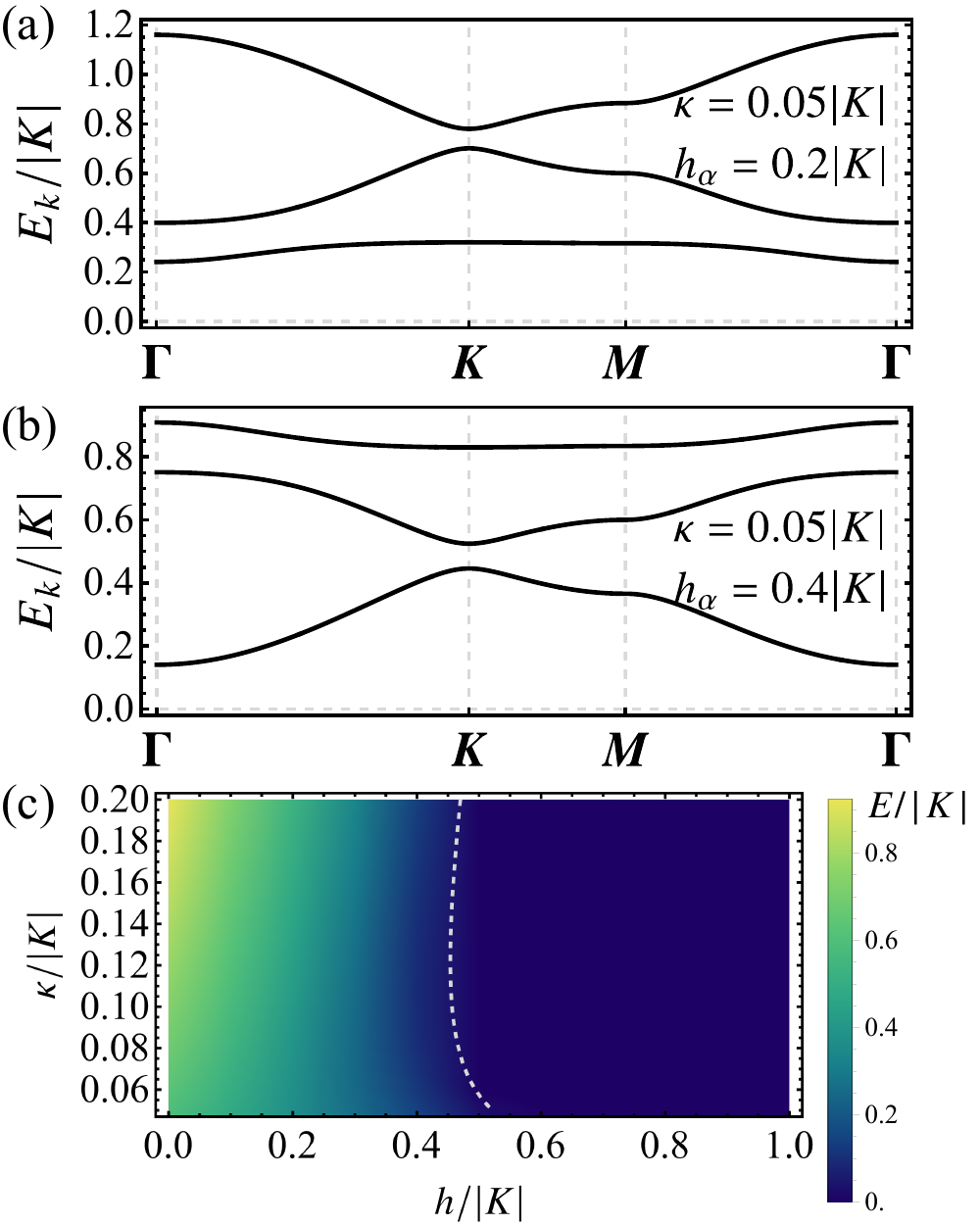}
\caption{
Bosonic vison pair bands in (a) FM and (b) AFM Kitaev models at selected $(\kappa,h)$ values.
In both cases, the band minimum is at the $\bGamma$ point.
(c) Boson band gap as a function of $h$ and $\kappa$ in the AFM Kitaev model.
The boson gap closes at the gray dashed line.
}\label{fig:b-bands}
\end{figure}

\subsection{Boson condensation and induced magnetic order}
\begin{figure}
\centering
\includegraphics[width=0.49 \textwidth]{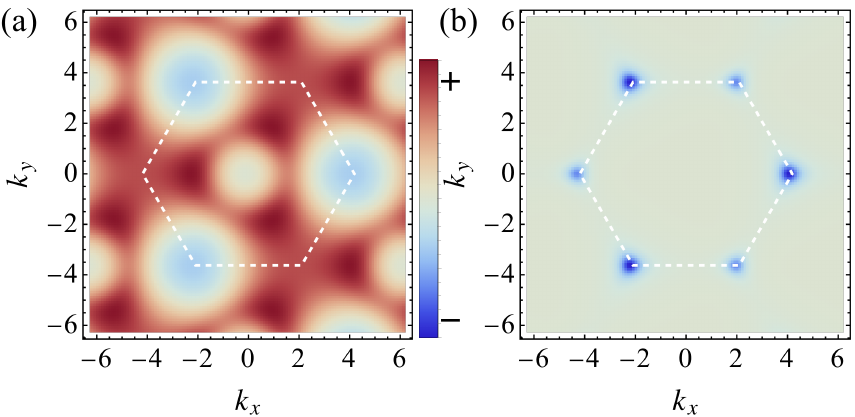}
\caption{
Berry curvature for the lowest boson band in (a) FM (Chern number $C = 1$) and 
(b) AFM (Chern number $C = -1$) Kitaev models.
The values of $(\kappa, h)$ are the same as those in \Fref{fig:b-bands}.
}\label{fig:b-BC}
\end{figure}

The bosonic vison pairs' band (from the hopping and on-site potential parts) for both FM and AFM Kitaev coupling at
selected $(\kappa, h)$ values are shown in \Fref{fig:b-bands}. 
As the $d_\mu$ bosons are moving on a Kagome lattice (see illustration in Fig.~2 of the main text),
there are three bands. The lowest energy band in both cases have non-trivial topology.
Its Chern number is $C = \pm 1$ for the FM/AFM Kitaev model. \Fref{fig:b-BC} shows its Berry curvature for each case.

As shown in \Frefs{fig:b-bands}(a)-(b), the band minimum is always at the $\bGamma$ point, i.e., the mode that become soft
has zero momentum in both FM and AFM cases.
In the AFM Kitaev model, because $\lambda_d = 0$, the lowest-energy mode (from the lowest-energy band $E_{\bk,1}$),
$\beta_{\bGamma,1}$, will condense at $h^\mathrm{AFM}_d$ when $E_{\bGamma,1} \rightarrow 0$. 
\Fref{fig:b-bands}(c) shows the band gap of bosons as a function of $h$ and $\kappa$.
It is interesting to notice that its wavefunction
has an orbital ``angular momentum'' $1 \, (\mathrm{mod} \ 3)$:
\begin{equation}
    \beta_{\bGamma,1}^\dagger = \frac{1}{\sqrt{3}} ( d_{\bGamma,x}^\dagger + e^{i2\pi/3} d_{\bGamma,y}^\dagger
    + e^{i4\pi/3} d_{\bGamma,z}^\dagger ).
\end{equation}
Therefore, in a $\beta_{\bGamma,1}$ condensed state:
$| \Phi \rangle \approx ( 1 + a e^{i\theta} \beta_{\bGamma,1}^\dagger + \dots )| \Omega \rangle$, the induced magnetic moment
at $\br \in A$ is:
\begin{align}
    \langle \bm{\sigma}_{\br} \rangle & \propto  ( \cos(\theta), \cos(\theta + 2\pi/3), \cos(\theta+4\pi/3) ) \nonumber \\ 
    & \perp (1,1,1).
\end{align}
Here we have chosen the gauge such that $\langle \Omega | \sigma_{\br}^\mu d_{\br,\mu}^\dagger | \Omega \rangle \in \mathbb{R}$
is independent of $\mu$ due to the $C_3$ rotational symmetry around site $\br$.
Because $d_{\br,\mu}^\dagger | \Omega \rangle$ is odd under spatial inversion around the center of bond $(\br,\mu)$,
as discussed in the previous section, there is:
$\langle \bm{\sigma}_{\br \in A} \rangle = -\langle \bm{\sigma}_{\br' \in B} \rangle$.
Therefore the condensation of $\beta_{\bGamma,1}$ in the AFM model implies an \emph{in-plane} N\'eel type magnetic order,
which seems to be aligned with recent observation of in-plane AFM-like magnetic modulation~\cite{Zhang2024-ec}.

In the FM Kitaev model, we have found that the lowest energy mode $\beta_{\bGamma,1}$ is an equal superposition
of all three types of bosons:
\begin{equation}
    \beta_{\bGamma,1}^\dagger = \frac{1}{\sqrt{3}} ( d_{\bGamma,x}^\dagger + d_{\bGamma,y}^\dagger + d_{\bGamma,z}^\dagger  ).
\end{equation}
as $\lambda_d \neq 0$, the single-particle creation and annihilation part reads:
\begin{align}
    \lambda_d \sum_{\br,\mu} (d_{\br,\mu} + d_{\br,\mu}^\dagger) = & \lambda_d  \sqrt{N} \sum_{\mu} (d_{\bGamma,\mu} + d_{\bGamma,\mu}^\dagger) \nonumber \\
    = & \lambda_d \sqrt{3N} ( \beta_{\bGamma,1} + \beta_{\bGamma,1}^\dagger ).
\end{align}
Combined with the band energy part: $E_{\bGamma,1} \beta_{\bGamma,1}^\dagger \beta_{\bGamma,1}$,
the part involving $\beta_{\bGamma,1}$ mode reads:
\begin{align}
    & \beta_{\bGamma,1}^\dagger \beta_{\bGamma,1} E_{\bGamma,1} + \sqrt{3N} \lambda_d ( \beta_{\bGamma,1} + \beta_{\bGamma,1}^\dagger ) \nonumber \\
    & = E_{\bGamma,1} \tilde{\beta}_{\bGamma,1}^\dagger \tilde{\beta}_{\bGamma,1} - 3N \lambda_d^2/E_{\bGamma,1},
\end{align}
with the bosonic mode $\tilde{\beta}_{\bGamma,1} = \beta_{\bGamma,1} + \sqrt{3N}\lambda_d/E_{\bGamma,1}$.
Therefore in the FM case the mode has trivial symmetry and transforms identically to the $[111]$ Zeeman field.
Therefore, there cannot be a sharp BEC-like transition associated with its condensation, 
but its softening is an indication of the continuous increase of the spin magnetization along the $[111]$ direction.
To see this more explicitly we consider a toy wavefunction in which there is a coherent occupation of the
$\beta_{\bGamma,1}$ mode:
\begin{equation} \label{eq:Phi-FM}
    | \Phi \rangle = e^{-3N (\lambda_d/E_{\bGamma,1})^2} e^{-\sqrt{3N}\frac{\lambda_d}{E_{\bGamma,1}}\beta_{\bGamma,1}^\dagger} | \Omega \rangle.
\end{equation}
Because $d_{\br,\mu}^\dagger | \Omega \rangle$ is even under inversion around the center of bond $(\br,\mu)$ (see the discussion before),
$| \Phi \rangle$ exhibits a uniform magnetic moment along the field direction.
Perturbatively in $\lambda_d \approx (-h) 2l$, with $l = \langle \Omega | \sigma_{\br}^\mu d_{\br,\mu}^\dagger | \Omega \rangle \in \mathbb{R}$,
there is:
\begin{align}
    \langle \bm{\sigma}_{\br} \rangle \approx  \frac{h}{E_{\bGamma,1}}4 l^2 (1,1,1).
\end{align}

\begin{figure}
\centering
\includegraphics[width=0.49 \textwidth]{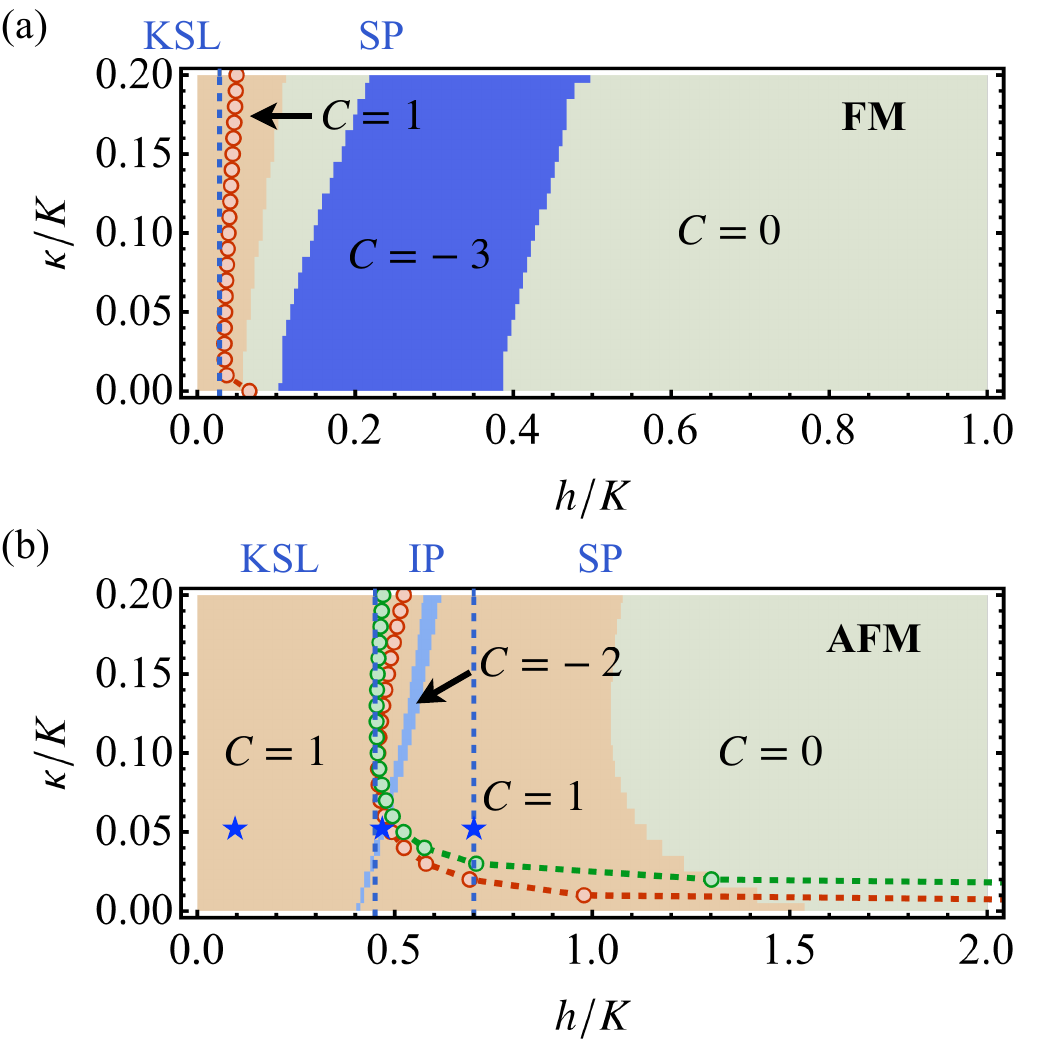}
\caption{
The phase diagram including small $\kappa$ regime.
Fermion Chern number (color plot) and the critical fields indicating proliferation of single visons
(red dotted line) and bosonic vison pairs (green dotted line)
for (a) FM and (b) AFM Kitaev couplings.
}\label{fig:full-phase-diagram}
\end{figure}
\section{Phase diagram including the small-$\kappa$ regime} \label{sec:pd-small-kappa}
In the main text, only the $\kappa \ge 0.05 |K|$ part of the AFM Kitaev coupling phase diagram was shown
to avoid the artifact that the single vison remains immobile to linear order in $h$ in the absence of $\kappa$
as discussed in Refs.~\cite{Chen2023-vo,Joy2022-ba}.
For completeness, here we present the phase diagram including also the small-$\kappa$ region in \Fref{fig:full-phase-diagram}.
Moreover, for FM Kitaev coupling, the fermion part actually also contains an intermediate $C = -3$ region
but which is located at much larger values of $h$ beyond the point where the vison gap closes,
and thus these phases are irrelevant to the discussion of the transition and not trustable since they happen at large $h$,
but we will present them here for the sake completeness.
Both the the small-$h$ $C=0$ to $C=-3$ transition and the high-field transition from $C=-3$ to $C=0$
phases are accompanied by a gap closing at the $\bGamma$ point, indicating that these two $C = 0$
phases are topologically equivalent (with the same Chern number and parity indices at the four high-symmetry points)
and are topologically trivial.
Since the single-vison gap already closes near the small-field $C = 1$ to $C = 0$ transition, suggesting 
the system subsequently enters a trivial SP state, the $C = -3$ regime should be regarded as an ``artifact''
of the perturbative fermion-only calculation.
To avoid confusion, we have not shown the $C = -3$ region explicitly in the phase diagram presented in the main text.

For the AFM Kitaev coupling, the leading-order-in-$h$ hoppings
for both single visons and bosonic vison pairs are quite
small when $\kappa \sim 0$, so the critical fields for their proliferation is rather large.
Since the third-order $\kappa$ term is generated self-consistently in the presence of a Zeeman field~\cite{Kitaev2006}, 
we believe the low-energy physics of single visons and bosonic vison pairs is best captured by a small, but finite, $\kappa$ value.
Note that the $\kappa$ term does not break any additional symmetry of the Kitaev plus Zeeman
model, and naturally regularizes the finite-size effect when extracting the 
single-vison hoppings~\cite{Chen2023-vo,Joy2022-ba}.
We note also that for the case of fermion-type quasiparticles, the Majorana gap at $\bK$ point at small $h$ is generated 
self-consistently by the three terms of the effective fermion model in Eq.~(4), and thus the inclusion of $\kappa$ is not that crucial
as it is for single visons and boson vison pairs.

\section{Ground state degeneracy of Kitaev model with a small $\kappa$ term} \label{sec:GS_parity}
\begin{figure}
\centering
\includegraphics[width=0.49 \textwidth]{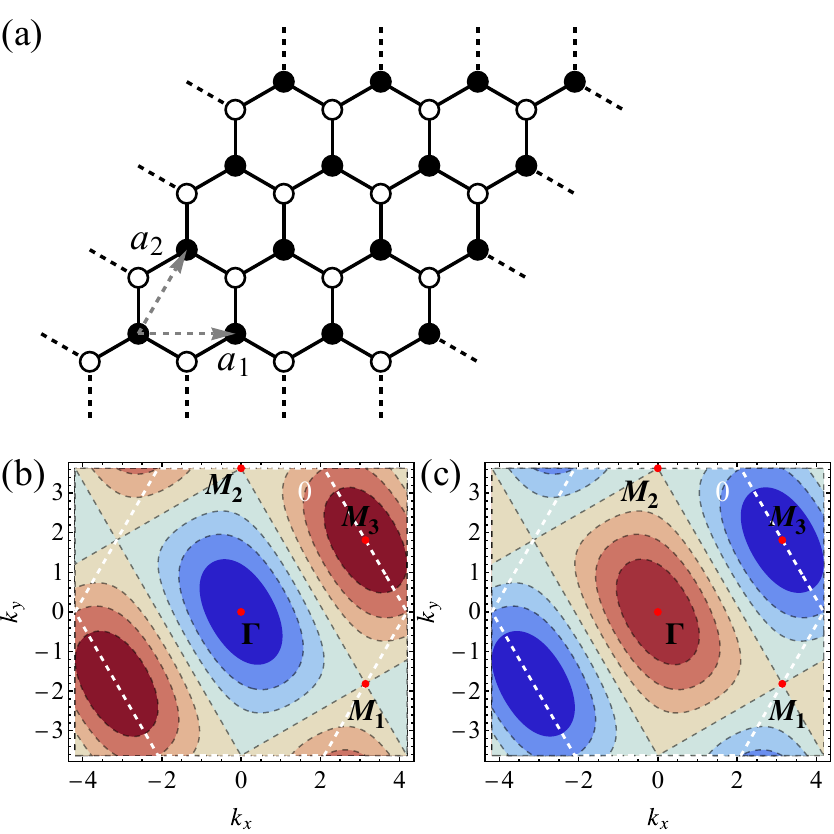}
\caption{
(a) Schematic of the periodic system on a torus ($N_1 = N_2 = 4$).
Matter fermions' energy dispersion in the zero-flux sector (with translational symmetry)
for FM (b) and AFM (c) Kitaev couplings.
}\label{fig:torus}
\end{figure}

In this section, we discuss the ground state degeneracy ($\mathcal{D}_\mathrm{GS}$)
of the $K+\kappa$ model (with $\kappa$ being small) on a torus.
The torus has $N_1 \times N_2$ unit cells, with $N_{1,2}$ standing for the linear size along 
each direction (see \Fref{fig:torus}(a) for an illustration).
We will prove that for the FM Kitaev model, 
$\mathcal{D}_\mathrm{GS} = 3$ independent of whether $N_\alpha$ is even or odd,
whereas for the AFM Kitaev model, there is a \emph{unique} ground state
when $(N_1, N_2)$ is (odd, odd), and
$\mathcal{D}_\mathrm{GS} = 3$ for other cases (see the summary in Table.~\ref{tab:D_GS}).
\begin{table}
    \centering
    \begin{tabular}{c | c c c c}
    \hline
    \hline
        & (e,~e) & (e,~o) & (o,~e) & (o,~o) \\
        \hline 
    FM  & 3 & 3 & 3 & 3 \\
    AFM & 3 & 3 & 3 & $\bm{1}$ \\
    \hline \hline
    \end{tabular}
    \caption{Summary of the ground state degeneracy for the $K+\kappa$ model ($\kappa$ being small) on a torus with different system sizes.
    Here, e (o) denotes even (odd).}
    \label{tab:D_GS}
\end{table}
%

For a periodic system, the local constraint of the Majorana representation of spins
$D_j = b_j^x b_j^y b_j^z c_j = 1$
indicates a global constraint  $\prod_j D_j = 1$ for physical states.
It can be shown that this can be recast into~\cite{Pedrocchi2011-re,Zschocke2015-zr}:
\begin{align} \label{eq:global-constr}
\prod_j D_j \nonumber & = (-1)^{N_1 + N_2} \prod_{\br \in A} ( -i c_{\br} c_{\br+\bdelta_x} )
\prod_{\br \in A, \alpha} ( -i b_{\br}^\alpha b_{\br+\bdelta_\alpha}^\alpha ) \\
& = 1.
\end{align}
One can define complex $a$-fermion modes from $c$-Majoranas within each unit cell:
\begin{align}
a_{\br} = \frac{1}{2} (c_{\br} + i c_{\br+\bdelta_x} ), \
a_{\br}^\dagger = \frac{1}{2} (c_{\br} - i c_{\br+\bdelta_x} ),\
\br \in A,
\end{align}
and bond $\chi_{\alpha}$-fermion from the $b^\alpha$ modes:
\begin{align}
\chi_{\br,\alpha} = \frac{1}{2} ( b_{\br}^\alpha + i b_{\br+\bdelta_\alpha}^\alpha ), \
\chi_{\br,\alpha}^\dagger = \frac{1}{2} ( b_{\br}^\alpha - i b_{\br+\bdelta_\alpha}^\alpha ), \
\br \in A.
\end{align}
There are thus $-i c_{\br} c_{\br+\bdelta_x} = (-1)^{ a_{\br}^\dagger a_{\br} }$, and
$ -i b_{\br}^\alpha b_{\br+\bdelta_\alpha}^\alpha = (-1)^{\chi_{\br,\alpha}^\dagger \chi_{\br,\alpha}}$.
\Eqref{eq:global-constr} thus sets a constraint of the total fermion parity (including both
the bond and matter fermions), depending on the system size.

The ground states of the $K+\kappa$ model (with small $\kappa$) are in the flux-free sector, and distinguished by
different Wilson loops of the $\bbZ_2$ gauge field, which sets the boundary condition for the matter
$a$-fermions.
In practice, one can choose the gauge $-i b_{\br}^\alpha b_{\br+\bdelta_\alpha}^\alpha = 1$, i.e., no $\chi_\alpha$ fermion,
for the case with periodic boundary conditions (PBC) along both directions, and only flip the sign of boundary terms for anti-periodic boundary conditions
(APBC). The BdG Hamiltonian of matter $a$-fermions under this gauge choice reads:
\begin{subequations}
\begin{align}
    H_a = & \frac{1}{2} \sum_k 
    \begin{pmatrix}
        a_k^\dagger & a_{-k}
    \end{pmatrix}
     h(k)
     \begin{pmatrix}
         a_k \\
         a_{-k}^\dagger
     \end{pmatrix}, \\
     h(k) = &
     \begin{pmatrix}
         \epsilon_k & \Delta_k \\
         \Delta_k^* & -\epsilon_k
     \end{pmatrix}.
\end{align}
\end{subequations}
Here
\begin{subequations}
\begin{align}
    \epsilon_k = & 2K [1+\cos(\bk \cdot \bm{a}_1) + \cos( \bk \cdot \bm{a}_2 )], \\
    \Delta_k = & -2i K [ \sin(\bk \cdot \bm{a}_1) + \sin(\bk \cdot \bm{a}_2) ] 
    + 4\kappa [ \sin(\bk \cdot \bm{a}_1) \nonumber \\
    & +\sin[\bk \cdot (\bm{a}_2-\bm{a}_1)] - \sin(\bk \cdot \bm{a}_2) ].
\end{align}
\end{subequations}
Below we analyze the $\mathcal{D}_\mathrm{GS}$ for FM and AFM Kitaev couplings respectively.

\subsection{$\mathcal{D}_\mathrm{GS}$ for the FM Kitaev model}
For FM Kitaev coupling, $K = -1$, the matter fermions' band dispersion at the four high-symmetry (unpaired) $\bk$-points
(HSPs) has the following property:
$\epsilon_{\bGamma}, \epsilon_{\bM_1}, \epsilon_{\bM_2} < 0 < \epsilon_{\bM_3}$ (see the plot in \Fref{fig:torus}(b)).
\begin{enumerate}
    \item (even,~even).
    In this case, $(-1)^{N_1+N_2} = 1$, and the constraint \Eqref{eq:global-constr} requires an even total fermion parity.
    \begin{enumerate}
        \item (PBC,~PBC).
        Since there is no $\chi_\alpha$-fermion, $(-1)^{\chi_{\br,\alpha}^\dagger \chi_{\br,\alpha}} = 1$, so physical states
        should have an even number of $a$-fermions. All the HSPs are included in the BdG Hamiltonian, and $\bGamma$, $\bM_1$, $\bM_2$
        modes are occupied, so the ground state of $H_a$ in this case is parity odd, thus unphysical.
        \item (PBC,~APBC).
        Because $N_1$ is even, so the parity of $\chi_\alpha$-fermions is even, and physical states should have an even $a$-parity.
        In this case, non of the HSPs are included in $H_a$, so its ground state has an even number of $a$-fermions, thus physical.
        \item (APBC,~PBC).
        The analysis is similar to the (PBC, APBC) case, and the ground state is physical.
        \item (APBC,~APBC).
        In this case, as both $N_1$ and $N_2$ are even, the total number of $\chi_\alpha$-fermions is even ($N_1 + N_2$), so physical
        states should have an even $a$-parity. Similar to previous case, non of the HSPs are included in $H_a$, so its ground state
        is parity even, thus physical.
    \end{enumerate}
    Therefore, the $\mathcal{D}_\mathrm{GS} = 3$ for an (even,~even) torus.
    \item (even,~odd).
    In this case, $(-1)^{N_1+N_2} = -1$, and the constraint \Eqref{eq:global-constr} requires an odd total fermion parity.
    \begin{enumerate}
        \item (PBC,~PBC)
        In this case, the $\chi_\alpha$-fermions' parity is even, so physical states should have an odd number of $a$-fermions.
        Only $\bGamma$ and $\bM_1$ are included in $H_a$, and both are occupied, thus the ground state of $H_a$ has an even number of $a$-fermions,
        thus unphysical.
        \item (PBC,~APBC)
        As $N_1$ is even, the total number of $\chi_\alpha$-fermions is even, and physical states should have an odd number of $a$-fermions.
        $\bM_2$ and $\bM_3$ points are included in $H_a$, and only $\bM_2$ are occupied, so the ground state of $H_a$ is parity odd, thus physical.
        \item (APBC,~PBC)
        As $N_2$ is odd, the total number of $\chi_\alpha$-fermions is odd, and physical states should have an even number of $a$-fermions.
        Non of the HSPs are included in $H_a$, so the ground state of $H_a$ is parity even, thus physical.
        \item (APBC,~APBC)
        The total number of $\chi_\alpha$-fermions is odd, and physical states should have an even numebr of $a$-fermions.
        Non of the HSPs are included in $H_a$, so the ground state of $H_a$ is parity even, thus physical.
    \end{enumerate}
    Therefore, the $\mathcal{D}_\mathrm{GS} = 3$ for an (even,~odd) torus.
    \item (odd,~even).
    The analysis of this case is similar to the (even,~odd) case, and $\mathcal{D}_\mathrm{GS} = 3$ for an (odd,~even) torus.
    \item (odd,~odd).
    In this case, $(-1)^{N_1+N_2}=1$, and physical states should have an even number of fermions.
    \begin{enumerate}
        \item (PBC,~PBC)
        The $\chi_\alpha$-fermions' number is even, so physical states should have an even $a$-parity. Only the $\bGamma$ point is included
        in $H_a$ and occupied, therefore the ground state of $H_a$ is parity odd, thus unphysical.
        \item (PBC,~APBC)
        The $\chi_\alpha$-fermions' parity is odd, so physical states should have an odd number of $a$-fermions.
        Only the $\bM_2$ point is included in $H_a$ and occupied, so the ground state of $H_a$ is parity odd, thus physical.
        \item (APBC,~PBC)
        The $\chi_\alpha$-fermions' number is odd, so physical states should have an odd number of $a$-fermions.
        Only the $\bM_1$ point is included in $H_a$ and occupied, so the ground state of $H_a$ is parity odd, thus physical.
        \item (APBC,~APBC)
        The $\chi_\alpha$-fermions' number is even, so physical states should have an even number of $a$-fermions.
        Only the $\bM_3$ point is indluded in $H_a$ but empty, so the ground state of $H_a$ is parity even, thus physical.
    \end{enumerate}
    Therefore, the $\mathcal{D}_\mathrm{GS} = 3$ for an (odd,~odd) torus.
\end{enumerate}

\subsection{$\mathcal{D}_\mathrm{GS}$ for the AFM Kitaev model}
For FM Kitaev coupling, $K = 1$, the matter fermions' band dispersion at the four high-symmetry (unpaired) $\bk$-points
(HSPs) has the following property:
$\epsilon_{\bGamma}, \epsilon_{\bM_1}, \epsilon_{\bM_2} > 0 > \epsilon_{\bM_3}$ (see the plot in \Fref{fig:torus}(c)).
\begin{enumerate}
    \item (even,~even).
    In this case, $(-1)^{N_1+N_2} = 1$, and the constraint \Eqref{eq:global-constr} requires an even total fermion parity.
    \begin{enumerate}
        \item (PBC,~PBC).
        The $\chi_\alpha$-fermions' number is even, so physical states should have an even number of $a$-fermions.
        All the HSPs are included in the BdG Hamiltonian, and only the $\bM_3$
        mode is occupied, so the ground state of $H_a$ is parity odd, thus unphysical.
        \item (PBC,~APBC).
        The parity of $\chi_\alpha$-fermions is even, and physical states should have an even $a$-parity.
        In this case, non of the HSPs are included in $H_a$, so its ground state has an even number of $a$-fermions, thus physical.
        \item (APBC,~PBC).
        The analysis is similar to the (PBC, APBC) case, and the ground state is physical.
        \item (APBC,~APBC).
        The total number of $\chi_\alpha$-fermions is even, so physical states should have an even $a$-parity.
        Similar to previous case, non of the HSPs are included in $H_a$, so its ground state is parity even, thus physical.
    \end{enumerate}
    Therefore, the $\mathcal{D}_\mathrm{GS} = 3$ for an (even,~even) torus.
    \item (even,~odd).
    In this case, $(-1)^{N_1+N_2} = -1$, and the constraint \Eqref{eq:global-constr} requires an odd total fermion parity.
    \begin{enumerate}
        \item (PBC,~PBC)
        In this case, the $\chi_\alpha$-fermions' parity is even, so physical states should have an odd number of $a$-fermions.
        Only $\bGamma$ and $\bM_1$ are included in $H_a$, and both are empty, thus the ground state of $H_a$ has an even number of $a$-fermions,
        thus unphysical.
        \item (PBC,~APBC)
        As $N_1$ is even, the total number of $\chi_\alpha$-fermions is even, and physical states should have an odd number of $a$-fermions.
        $\bM_2$ and $\bM_3$ points are included in $H_a$, and only $\bM_3$ are occupied, so the ground state of $H_a$ is parity odd, thus physical.
        \item (APBC,~PBC)
        As $N_2$ is odd, the total number of $\chi_\alpha$-fermions is odd, and physical states should have an even number of $a$-fermions.
        Non of the HSPs are included in $H_a$, so the ground state of $H_a$ is parity even, thus physical.
        \item (APBC,~APBC)
        The total number of $\chi_\alpha$-fermions is odd, and physical states should have an even numebr of $a$-fermions.
        Non of the HSPs are included in $H_a$, so the ground state of $H_a$ is parity even, thus physical.
    \end{enumerate}
    Therefore, the $\mathcal{D}_\mathrm{GS} = 3$ for an (even,~odd) torus.
    \item (odd,~even).
    The analysis of this case is similar to the (even,~odd) case, and $\mathcal{D}_\mathrm{GS} = 3$ for an (odd,~even) torus.
    \item (odd,~odd).
    In this case, $(-1)^{N_1+N_2}=1$, and physical states should have an even number of fermions.
    \begin{enumerate}
        \item (PBC,~PBC)
        The $\chi_\alpha$-fermions' number is even, so physical states should have an even $a$-parity. Only the $\bGamma$ point is included
        in $H_a$ but empty, therefore the ground state of $H_a$ is parity even, thus physical.
        \item (PBC,~APBC)
        The $\chi_\alpha$-fermions' parity is odd, so physical states should have an odd number of $a$-fermions.
        Only the $\bM_2$ point is included in $H_a$ and empty, so the ground state of $H_a$ is parity even, thus unphysical.
        \item (APBC,~PBC)
        The $\chi_\alpha$-fermions' number is odd, so physical states should have an odd number of $a$-fermions.
        Only the $\bM_1$ point is included in $H_a$ and empty, so the ground state of $H_a$ is parity even, thus unphysical.
        \item (APBC,~APBC)
        The $\chi_\alpha$-fermions' number is even, so physical states should have an even number of $a$-fermions.
        Only the $\bM_3$ point is indluded in $H_a$ and occupied, so the ground state of $H_a$ is parity odd, thus unphysical.
    \end{enumerate}
    Therefore, the $\mathcal{D}_\mathrm{GS} = 1$ for an (odd,~odd) torus.
\end{enumerate}

\bibliography{reference.bib}


\end{document}